\newcommand*{\ATLASLATEXPATH}{}
\documentclass[texlive=2016, UKenglish,cernpreprint]{\ATLASLATEXPATH atlasdoc}
 
\makeatletter
\DeclareOldFontCommand{\rm}{\normalfont\rmfamily}{\mathrm}
\DeclareOldFontCommand{\sf}{\normalfont\sffamily}{\mathsf}
\DeclareOldFontCommand{\tt}{\normalfont\ttfamily}{\mathtt}
\DeclareOldFontCommand{\bf}{\normalfont\bfseries}{\mathbf}
\DeclareOldFontCommand{\it}{\normalfont\itshape}{\mathit}
\DeclareOldFontCommand{\sl}{\normalfont\slshape}{\@nomath\sl}
\DeclareOldFontCommand{\sc}{\normalfont\scshape}{\@nomath\sc}
\makeatother

\usepackage{\ATLASLATEXPATH atlaspackage}
\usepackage[export]{adjustbox}
\usepackage{multirow}
\usepackage{floatrow}
\usepackage{subfig}
\usepackage{\ATLASLATEXPATH atlasbiblatex}
 
\usepackage{\ATLASLATEXPATH atlasphysics}
 
\graphicspath{{logos/}{figures/}}
 
\usepackage{ANA-HDBS-2018-58-PAPER-defs}

\def\bbar {\ensuremath{{\overline b}}}

\def\bbbar {\ensuremath{{b\bbar}}}

\def\invfb {\ensuremath{{\mathrm{fb}^{-1}}}}

\def\fb {\ensuremath{{\mathrm{fb}}}}
\def\pb {\ensuremath{{\mathrm{pb}}}}
\def\nb {\ensuremath{{\mathrm{nb}}}}

\def\tauhad{\ensuremath{\tau_{\text{had}}}}
\def\taulep{\ensuremath{\tau_{\text{lep}}}}

\newcommand{\LUMI}{36.1\,\invfb }
 
\usepackage{siunitx}
\usepackage{float}
\floatstyle{plaintop}
\restylefloat{table}

\AtlasTitle{Combination of searches for Higgs boson pairs in $pp$
collisions at $\sqrt{s} = $13~\TeV\
with the ATLAS detector}
 
\AtlasAbstract{
This letter presents a combination of searches for Higgs boson pair production using up to \LUMI\ of
proton--proton collision data at a centre-of-mass energy $\sqrt{s} = 13$~\TeV\ recorded with the
ATLAS detector at the LHC. The combination is performed using six analyses
searching for Higgs boson pairs decaying into
the \bbbb, \bbww, \bbtautau, \wwww, \bbyy and \wwyy final states.
Results are presented for non-resonant and resonant
Higgs boson pair production modes.
No statistically significant excess in data above the Standard Model predictions is found.
The combined observed (expected) limit at 95\% confidence level on the
non-resonant Higgs boson pair production cross-section is $\lorSMObsFull$ ($\lorSMExpFull$) times the predicted Standard Model
cross-section. Limits are also set on the ratio ($\kappa_{\lambda}$) of the Higgs boson
self-coupling  to its Standard Model value. This ratio is constrained
at 95\% confidence level in observation (expectation) to
$\limitLambdaOBSlow < \kappa_{\lambda} <\limitLambdaOBShigh$
($\limitLambdaEXPlow < \kappa_{\lambda} <\limitLambdaEXPhigh$).
In addition, limits are set on the production of
narrow scalar resonances and spin-2 Kaluza--Klein Randall--Sundrum
gravitons. Exclusion regions are also provided in the parameter space of the
habemus Minimal Supersymmetric Standard Model and the Electroweak Singlet
Model.
}
 
\author{The ATLAS Collaboration}
 
\AtlasRefCode{HDBS-2018-58}

\AtlasJournal{Phys. Lett. B}
 
\PreprintIdNumber{CERN-EP-2019-099}
 
\AtlasCoverSupportingNote{The combination and interpretation of $HH$ searches in $pp$ collisions at $\sqrt{s} = 13$~TeV with the ATLAS detector}{https://cds.cern.ch/record/2653800/}
 
\AtlasCoverCommentsDeadline{}
 
\AtlasCoverAnalysisTeam{
John Alison,
Nansi Andari,
Florian Beisiegel,
Alessandra Betti,
Petar Bokan,
Elizabeth Brost,
Patrick Bryant,
Tyler Burch,
Leo Cerda,
David Delgove,
Biagio Di Micco,
Jochen Dingfelder,
David Englert,
Yaquan Fang,
Arnaud Ferrari,
Douglas Gingrich,
Kathryn Grimm,
Carl Gwilliam,
Jonathan Hays,
Tatjana Lenz,
Qi Li,
Gabriel Palacino,
James Robinson,
Nikolaos Rompotis,
Eleonora Rossi,
Jana Schaarschmidt,
Chase Shimmin,
Magdalena Slawinska,
Xiaohu Sun,
Alan Taylor,
Tulin Varol,
David Wardrope,
Yu Zhang
}
 
\AtlasCoverEdBoardMember{Kathrin Becker, John Hobbs, Jianming Qian~(chair)}

\AtlasCoverEgroupEditors{atlas-hdbs-2018-58-editors@cern.ch}
 
\AtlasCoverEgroupEdBoard{atlas-hdbs-2018-58-editorial-board@cern.ch}
 
\AtlasJournalRef{Phys. Lett. B 800 (2020) 135103}
\AtlasDOI{10.1016/j.physletb.2019.135103}
 
\hypersetup{pdftitle={ATLAS document},pdfauthor={The ATLAS Collaboration}}
 
\begin{document}
 
\maketitle

\section{Introduction}
\label{sec:intro}
 
The discovery of the Higgs boson ($H$)~\cite{Aad:2012tfa,Chatrchyan:2012xdj} at the Large Hadron Collider (LHC)~\cite{Evans:2008} in
2012 has experimentally confirmed the Brout--Englert--Higgs (BEH) mechanism of electroweak symmetry breaking and
mass generation~\cite{Englert:1964et,Higgs:1964pj,Guralnik:1964eu}. The BEH mechanism not only predicts the existence
of a massive scalar particle, but also requires this scalar particle to couple to itself. Therefore,
observing the production of Higgs boson pairs ($HH$) and measuring the Higgs boson self-coupling $\lambda_{HHH}$
is a crucial validation of the BEH mechanism. Any deviation from the Standard Model (SM) predictions
would open a window to new physics. Moreover, the form of the Higgs field
potential, which generates the Higgs boson self-coupling after electroweak symmetry breaking, can
have important cosmological implications, involving, for example, predictions for vacuum stability or models in which the
Higgs boson acts as the inflation field \cite{Bezrukov:2007ep,
Masina:2011aa, Branchina:2008pc, Buttazzo:2013uya}.
 
\begin{figure}[htbp]
\centering
\subfloat[]{\includegraphics[width=0.40\textwidth,valign=c]{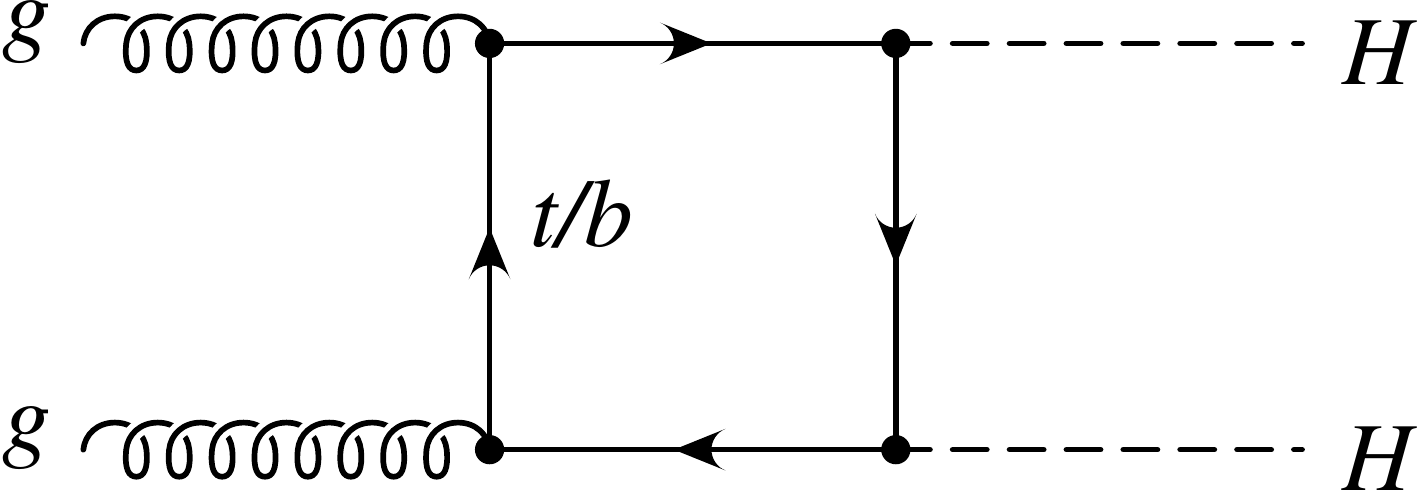}\label{fig:hh-diagrams_a}}
\quad \quad
\subfloat[]{\includegraphics[width=0.40\textwidth,valign=c]{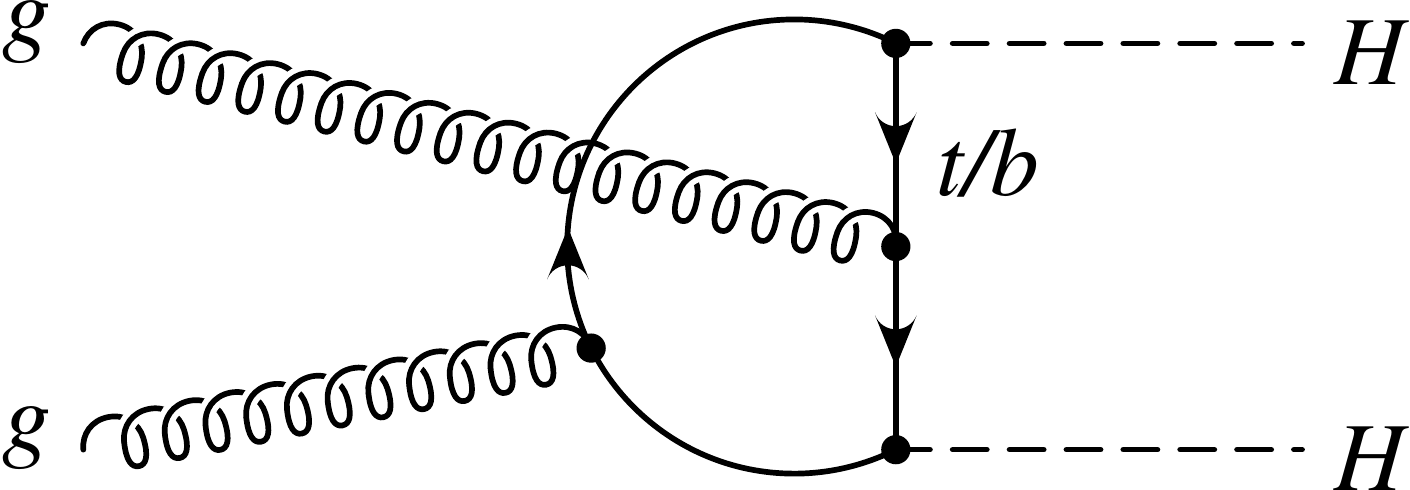}\label{fig:hh-diagrams_b}} \\ 
\subfloat[]{\includegraphics[width=0.40\textwidth,valign=c]{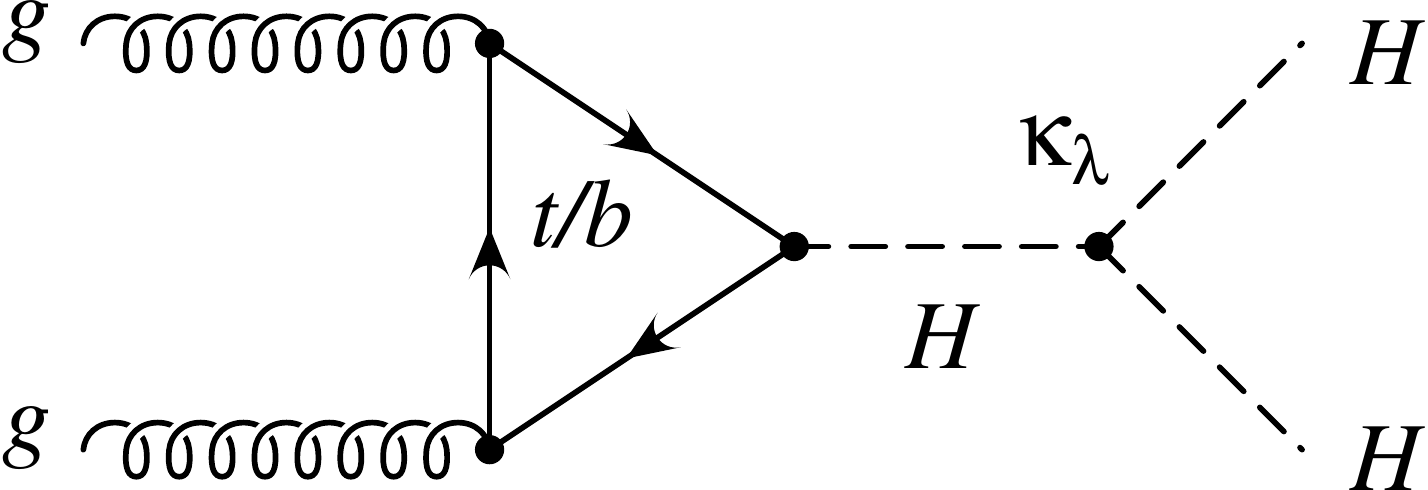}\label{fig:hh-diagrams_c}}
\quad \quad \quad \quad
\subfloat[]{\includegraphics[width=0.33\textwidth,valign=c]{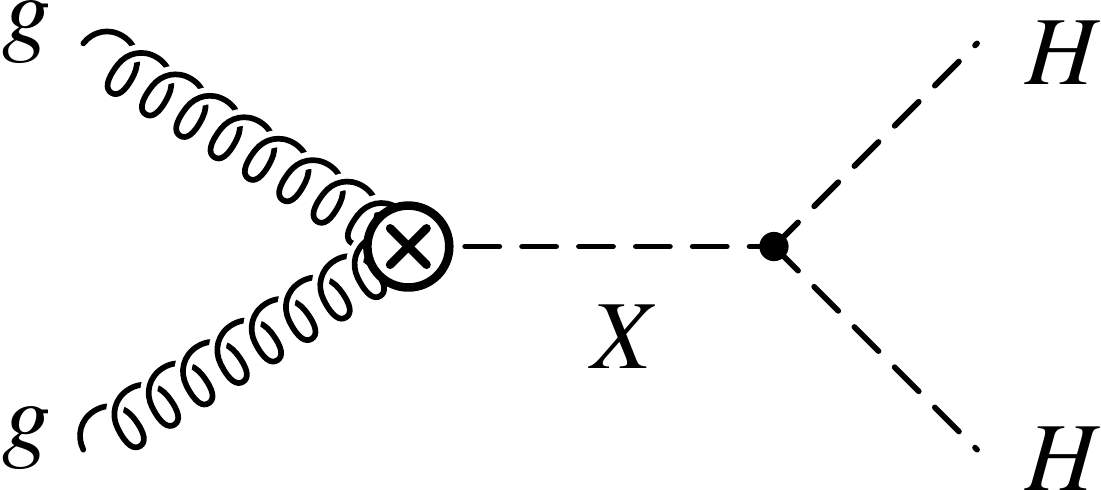}\label{fig:hh-BSMdiagrams_d}}

\caption{\label{fig:hh-diagrams} Examples of leading-order Feynman
diagrams for Higgs boson pair production: the diagrams
(a) and (b) are proportional to the square of the heavy-quark
Yukawa couplings, while the diagram (c) is proportional to the product
of the heavy-quark Yukawa coupling and the Higgs boson
self-coupling. Here $\kappa_{\lambda}$ is the ratio of the beyond-the-Standard-Model Higgs boson self-coupling to that of the SM. The diagram (d) represents the production of the Higgs boson pair through
an intermediate resonance ($X$) that couples to gluons through an
effective coupling  and to the SM Higgs boson.}
\end{figure}

In the SM, the gluon--gluon fusion $pp \to HH$ process (ggF) accounts for more than 90\% of the Higgs boson pair
production cross-section, and only this production mode is considered here.
It proceeds via two amplitudes: the first ($\mathcal{A}_1$) represented by the diagrams
(a) and (b), and the second ($\mathcal{A}_2$) represented by the diagram (c) in Figure~\ref{fig:hh-diagrams}.
The interference between these two amplitudes is destructive and
yields an overall cross-section of $\sigma^{\rm SM}_{\rm ggF} (pp \to HH) = 33.5^{+2.4}_{-2.8}$~fb at
$\sqrt{s} = 13$~\TeV~\cite{deFlorian:2016spz}, calculated first  at
next-to-leading order (NLO) in QCD with the
heavy top-quark approximation~\cite{Dawson:1998py}, then numerically with full top-quark mass
dependence~\cite{Borowka:2016ehy} (confirmed later in
Ref.~\cite{Baglio:2018lrj} and analytically computed with some approximation in Ref.~\cite{Bonciani:2018omm}) corrected at next-to-next-to-leading order
(NNLO) \cite{deFlorian:2013jea} in QCD matched with
next-to-next-to-leading logarithmic (NNLL)
resummation in the heavy top-quark
limit~\cite{Shao:2013bz,deFlorian:2015moa}. The Higgs boson mass used in these calculations and for all results in this paper is $m_{H} =$ 125.09 GeV~\cite{HIGG-2014-14}.
Beyond-the-Standard-Model (BSM) scenarios can bring substantial
enhancement of this cross-section by modifying the relative sign
of $\mathcal{A}_1$ and $\mathcal{A}_2$, and by increasing $\mathcal{A}_2$.
The $\mathcal{A}_2$ amplitude is proportional to the Higgs self-coupling  $\lambda_{HHH}$.
The Higgs boson self-coupling modifier due to BSM scenarios is defined as
$\kappa_{\lambda} = \lambda_{HHH}/\lambda^{\rm
SM}_{HHH}$. In this analysis, all other Higgs boson couplings are assumed to have SM values.
Indirect limits on $\kappa_{\lambda}$ have been obtained
using the measurements of single Higgs boson production and decay~\cite{Degrassi:2016wml} and
electroweak precision observables~\cite{Degrassi:2017ucl,
Kribs:2017znd}, constraining  $\kappa_{\lambda}$ to the range $-8 <
\kappa_{\lambda} < 14$ at 95\% confidence level (CL).
The Higgs boson self-coupling is discussed in the context of BSM models
in Refs.~\cite{Kribs:2017znd, DiLuzio:2017tfn}.
 
Several BSM models also predict the existence of heavy particles decaying into a pair of Higgs bosons.
Two-Higgs-Doublet Models \cite{Branco:2011iw}, models inspired by the
Minimal Supersymmetric Standard Model (MSSM) like habemus MSSM (hMSSM) ~\cite{Djouadi:2005gj,Djouadi:2013vqa,Djouadi:2013uqa,Djouadi:2015jea},
and Electroweak Singlet Models
(EWK-singlet)~\cite{Hill:1987ea,Veltman:1989vw,Robens:2015gla,deFlorian:2016spz}
predict, in addition to the Higgs boson,  a second, heavier, CP-even scalar that can decay into two SM Higgs bosons.
In the EWK-singlet model, the scalar states are mixed, with a mixing angle $\alpha$.
The ratio of the vacuum expectation value of the additional singlet
to that of the SM Higgs doublet, $\tan\beta$, is a free parameter.
In the hMSSM, the CP-even states also mix, and the model's phenomenology can be described by the mass ($m_A$) of a third, CP-odd,
resonance and the ratio of the vacuum expectation values of the two Higgs doublets, $\tan\beta$.
Alternatively, the Higgs boson pair can be produced resonantly through
the decay of a spin-2 Kaluza--Klein (KK) graviton,
as predicted in the Randall--Sundrum (RS) model of warped extra
dimensions~\cite{Randall:1999ee}. A schematic diagram for production of a heavy
resonance followed by its decay into a Higgs boson
pair is shown in Figure \ref{fig:hh-BSMdiagrams_d}.
 
This letter presents a combination of results from searches for both non-resonant and resonant
Higgs boson pair production in proton--proton ($pp$) collisions at
$\sqrt{s} = 13$~\TeV. The data were collected with the
ATLAS detector~\cite{PERF-2007-01,ATLAS-TDR-19, Abbott:2018ikt}  and correspond to an integrated luminosity of up to \LUMI.
The combination includes all published ATLAS $HH$ search analyses
using $\sqrt{s}=13$ TeV data, namely those studying the final states
\bbbb~\cite{Aaboud:2018knk}, \bbww~\cite{Aaboud:2018zhh},
\bbtautau~\cite{Aaboud:2018sfw}, \wwww~\cite{Aaboud:2018ksn},
\bbyy~\cite{Aaboud:2018ftw} and \wwyy~\cite{Aaboud:2018ewm}.
 
Previous combinations of searches for $HH$ pair production were
performed at $\sqrt{s} = 8$ TeV by the ATLAS experiment
\cite{Aad:2015xja} and at $\sqrt{s} = 13$ TeV by the CMS experiment
\cite{Sirunyan:2018two} combining the final states \bbbb
\cite{Sirunyan:2018zkk,Sirunyan:2017isc,Sirunyan:2018qca,Sirunyan:2018tki},
$bb$VV \cite{Sirunyan:2017guj},
\bbtt \cite{Sirunyan:2017djm} and \bbyy \cite{Sirunyan:2018iwt}.
 
\section{Analysis description}
\label{ana_descr}
 
The analysis strategies for each of the final states considered in this letter are summarised below.
\begin{itemize}
\item The \bbbb analysis is performed using four anti-$k_t$ jets reconstructed
with a radius parameter $R= 0.4$~\cite{Cacciari:2008gp, Fastjet}
(resolved analysis) or two large-$R$ jets with $R = 1.0$ (boosted analysis). The
dataset of the resolved  analysis is split according to the years 2015 and 2016,  and then statistically
combined taking into account the different trigger algorithms used in  2015 and
2016. In part of the 2016 data period, inefficiencies in
the online vertex reconstruction affected $b$-jet triggers that were
used in the resolved analysis, reducing the total available integrated luminosity  to 27.5
fb$^{-1}$. The boosted analysis searches for two
large-$R$ jets containing the $b$-quark pairs from the decays of the two Higgs bosons. The large-$R$ jets are
identified as originating from $b$-quarks using a
$b$-tagging algorithm applied to $R=0.2$ track-jets~\cite{ATL-PHYS-PUB-2014-013} associated
with the large-$R$ jet~\cite{Cacciari:2007fd}. The analysis is divided into  three categories: the first category  selects events in which each of
the two large-$R$ jets has one $b$-tagged track-jet; the second category requires that
one large-$R$  jet contains two $b$-tagged track-jets and
the other large-$R$  jet contains one $b$-tagged track-jet; the third
category requires that  both large-$R$ jets  contain two $b$-tagged track-jets.
For the SM $HH$ search, only the resolved analysis is used, with two categories,
one for the 2015 and another for the 2016 dataset. The resonant
search is instead performed with the resolved analysis for
masses in the range 260--1400 GeV, with the boosted analysis for masses in the
range 800--3000 GeV, and with the combination of the two for masses in the overlapping range
800--1400 GeV.
\item The \bbww analysis looks for the $WW \to \ell\nu qq$ decay channel,
where $\ell$ is an electron or muon,  and $q$ is a $u,d,s,c$ quark or
anti-quark.  The $b\bar{b}$ pair is selected from two $R = 0.4$ jets (resolved analysis)  or one $R = 1.0$  large-$R$
jet (boosted analysis), while the jets from the $W$ decay are
reconstructed with $R = 0.4$ jets. The resolved analysis is used
in the SM $HH$ search,  in the search for a scalar resonance with a mass between 500
and 1400 GeV, and in the search for a KK graviton in the mass range
500 to 800 GeV. The boosted analysis looks for scalar resonances in the
mass range 1400 to 3000 GeV and for KK gravitons between 800 and
3000 GeV. The resolved and boosted analyses each use one category. The two analyses are not
statistically combined due to a significant overlap between the two
signal regions.
 
\item The \bbtt analysis looks for final states with two
$R=0.4$ $b$-tagged jets and two $\tau$-leptons. One of the two $\tau$-leptons of the
$\tau^+ \tau^-$ pair is required to decay hadronically, while the other
decays either hadronically ($\tau_{\rm had} \tau_{\rm had}$) or
leptonically ($\tau_{\rm lep}\tau_{\rm had}$). In the $\tau_{\rm
lep} \tau_{\rm had}$ channel, events are
triggered by single lepton triggers (SLT), requiring  an electron or
a muon in the final state, or by the coincidence of a lepton
trigger with a hadronic $\tau$ trigger (LTT). In the $\tau_{\rm had}  \tau_{\rm had}$
channel, events are triggered by single
hadronic $\tau$ triggers (STT) or double hadronic $\tau$ triggers
(DTT). The analysis is divided into three categories: one selects $\tau_{\rm
had}\tau_{\rm had}$ events, a second selects $\tau_{\rm lep}\tau_{\rm
had}$ events triggered by the SLT, and a third  selects $\tau_{\rm lep}\tau_{\rm
had}$ events triggered by the LTT. The $\tau_{\rm had}\tau_{\rm had}$ and
the SLT $\tau_{\rm lep}\tau_{\rm had}$ categories are used for all
model interpretations, while the LTT $\tau_{\rm lep}\tau_{\rm had}$ category
is used in the SM $HH$ search (excluding the
$\kappa_{\lambda}$ analysis) and in resonant searches up
to a mass of 800 GeV.
 
\item The \wwww analysis looks for channels with leptonic and/or
hadronic $W$ decays. Three channels are identified: $\ell \nu \,
\ell\nu \, 4q$, $\ell \nu \, \ell \nu \, \ell \nu \, 2q$, and $\ell
\nu \, \ell \nu \, \ell \nu \, \ell \nu$, with $\ell$ being an
electron or muon, $q$ a quark, and $\nu$ a neutrino. The $q$ momentum is reconstructed from $R =
0.4$ jets. In order to suppress $Z+$jets and $t\bar{t}$ background, dilepton events are required to have two
leptons of the same charge. Events are categorised
according to the lepton flavour ($ee$, $e\mu$ and $\mu \mu$).  Three-lepton events are selected if the
sum of the lepton charges is $\pm$1. They are divided into two categories according to the number
of same-flavour, opposite-charge (SFOS) lepton pairs; one category
selects zero SFOS lepton pairs and a second category selects one or two SFOS
lepton pairs. Four-lepton
events are categorised according to the number of SFOS lepton pairs and the
invariant mass ($m_{4\ell}$) of the four-lepton system. Four
categories are defined, requiring that the number of SFOS lepton pairs is less than two or
equal to two, and $m_{4\ell}$ is smaller or larger than 180 GeV. A total of
nine categories are fit simultaneously in the searches for both non-resonant and resonant $HH$ production.
 
\item The \bbyy analysis searches for a $HH$ pair decaying into
$b\bar{b}$ and $\gamma \gamma$.
Two high-\pt~isolated photons are required to have $\ET/m_{\gamma\gamma} > $ 0.35 and 0.25 respectively.
The events are then analysed using two selections: a `loose selection' requiring  a jet with   $\pt > 40$ GeV
and a second jet with $\pt > 25$ GeV, and a `tight selection' where the two jets are required
to have $\pt$ larger than 100 and 30 GeV. All jets have a radius parameter
$R = 0.4$.
Both selections are subdivided into two categories requiring one
$b$-tagged jet or two $b$-tagged jets. The tight selection is used in the SM $HH$ search
and in the search for resonances with masses higher than 500 GeV,
while the loose selection is used in the $\kappa_{\lambda}$ analysis
and in the search for resonances  with masses smaller than 500 GeV.
The analysis is therefore divided into four categories, but only two of
them are simultaneously fit to extract each result.
 
\item The \wwyy analysis searches for a $HH$ pair decaying into  $\gamma  \gamma$
and  $WW$. The analysis uses the same photon selection as the \bbyy channel and looks for one $W$ decaying
leptonically and a second $W$ decaying hadronically ($WW \to \ell\nu  qq$).
The hadronic $W$ decay is reconstructed from $R=0.4$ jets. Only one category
is used in the searches for both non-resonant and resonant $HH$ production.

\end{itemize}

A summary of the main analysis characteristics is given in Table~\ref{tab:discriminators}.
All analyses impose a series of sequential  requirements on kinematic variables to select
signal events and suppress backgrounds. The \bbtt analysis uses a boosted decision tree
(BDT)~\cite{TMVA}  distribution as the final discriminant for both the non-resonant and resonant searches.
For the resonant searches, the \bbbb, \bbww and \bbyy analyses use the $HH$ invariant
mass ($m_{HH}$) as the final discriminant, the \wwyy analysis uses
the $\gamma \gamma$ invariant mass ($m_{\gamma \gamma}$), while \wwww
uses simple event counting.
For the SM $HH$ search, the \bbbb analysis uses the $m_{HH}$
distribution as a discriminant, profiting from the difference between the shapes of
the signal and the dominant multi-jet background. The \bbyy and
\wwyy analyses fit the $m_{\gamma \gamma}$ distribution to extract
both the signal yield  and the background expectation,
while the \bbww and \wwww analyses use event counting.
 
\begin{table}
{\def\arraystretch{1.1} \setlength{\tabcolsep}{4pt}
\begin{tabular}{l|c|c|c|c|c|c}
\footnotesize
& \bbbb & \bbww & \bbtautau & {\small \wwww} & \bbyy & \wwyy \\
\hline
$\mathcal{B}(HH \to x\bar{x}y\bar{y})$ & 0.34 & 0.25 & 0.073 & 0.046
& $2.6\cdot 10^{-3}$ &
$1.0\cdot
10^{-3}$
\\
$\mathcal{L}_{\rm int}$ [fb$^{-1}$] & 27.5 [36.1] & 36.1 & 36.1 &
36.1
& 36.1 & 36.1 \\
Categories & 2 [2--5] & 1 [1] & 3 [2--3] & 9 [9] & 2 [2] & 1 [1] \\
Discriminant  & $m_{HH}$ [$m_{HH}$]&  c.e.  [$m_{HH}$] & BDT [BDT]
& c.e. [c.e.] &  $m_{\rm \gamma \gamma}$ [$m_{HH}$] &
$m_{\rm \gamma
\gamma}$
[$m_{\rm \gamma \gamma}$] \\
Model & NR [$S/G$]  & NR [$S/G$]  & NR [$S/G$]  & NR [$S$] & NR [$S$] & NR [$S$] \\
$m_{S/G}$  [TeV] & [0.26--3.00] & [0.50--3.00] & [0.26--1.00] & [0.26--0.50] &
[0.26--1.00]&
[0.26--0.50]
\\
\hline
Ref. & \cite{Aaboud:2018knk} & \cite{Aaboud:2018zhh} &
\cite{Aaboud:2018sfw}
& \cite{Aaboud:2018ksn} &
\cite{Aaboud:2018ftw}
&
\cite{Aaboud:2018ewm} \\
\end{tabular}
}
\caption{Summary of the main characteristics of the analyses used in the non-resonant and resonant searches. The resonant analysis
characteristics are indicated between square brackets.  ``$\mathcal{B}(HH\to
x\bar{x}y\bar{y})$''   indicates the branching fraction of the $HH$ pair where one $H$ decays into
$x\bar{x}$ and the other decays into $y\bar{y}$.
The branching fraction values are taken from
Ref.~\cite{deFlorian:2016spz} for a Higgs boson mass
$m_H = 125.09$ GeV. ``$\mathcal{L}_{\rm
int}$'' indicates the integrated luminosity of the dataset used
in the analysis.  ``Categories''  indicates the number of signal
categories. ``Discriminant''  indicates the distribution  used in the final limit-setting fit
(``c.e.'' stands for counting events and indicates that a simple
event counting was used in the final fit rather than a distribution shape).
``Model''  indicates which models each analysis tested: NR stands for
SM $HH$ signal model, $S$ for a spin-0 scalar model, and  $G$ for a KK graviton model.
``$m_{S/G}$'' gives the probed mass range for the resonant
search. ``Ref.'' reports the reference to the individual final state papers. } \label{tab:discriminators}
\end{table}
 
\section{Statistical treatment}
The statistical interpretation of the combined search results is based on a simultaneous
fit to the data for the cross-section of the signal process and  nuisance parameters that encode
statistical and systematic uncertainties, using the CL$_{\rm S}$
approach~\cite{Read:2002hq}.
The asymptotic approximation~\cite{Cowan:2010js} is used in the
analysis of all final states and their combination.

All signal regions considered in the simultaneous fit are either orthogonal by construction or have
negligible overlap.
The overlap due to object misidentification between \bbyy and \bbtt,
and between \wwyy and \wwww, which are not orthogonal by construction,
is evaluated by running the signal and data samples from each channel through the analysis selection of each other channel.
Less than 0.1\% of simulated signal events overlap between analyses, and no overlap is found in data.
There is some irreducible contamination from \bbww and \bbZZ events
with $\tau$'s in the final state passing the \bbtautau selection.
This contamination is less than 8\% of the \bbtt selected events, and
it is not taken into account in the \bbtautau analysis, note that
including this contribution would increase the analysis sensitivity
therefore the results obtained here are slightly conservative.
The detector systematic uncertainties, such as
those in jet reconstruction, $b$-jet tagging, electron, muon and photon
reconstruction and identification, as well as the uncertainty on the
integrated luminosity~\cite{Aaboud:2016hhf}, are correlated across all
final states. Uncertainties on the signal acceptance derived
by varying the renormalisation and factorisation scales, the parton
distribution functions and the parton shower are correlated too.
Theoretical and modelling systematic uncertainties of the backgrounds
derived using simulated events are not correlated across different analyses
because the  overlap among their contributions to the different analyses is negligible.
 
\section{Combination of results on non-resonant Higgs boson pair production}
\label{sec:nonres}
 
The SM $HH$ analyses use signal samples generated at next-to-leading order (NLO) in QCD
with \textsc{Madgraph5\_aMC@NLO}~\cite{Alwall:2014hca}
using the CT10 NLO parton distribution function (PDF) set
\cite{Lai:2010vv}. Parton showers and hadronisation were simulated
with \textsc{Herwig++}~\cite{Bahr:2008pv} using parameter values
from the \textsc{UE-EE-5-CTEQ6L1} tune~\cite{Seymour:2013qka}.
The so-called FTApprox method~\cite{Frederix:2014hta}
is applied in the event generation to include finite top-quark mass effects in the real-radiation NLO corrections.
The virtual-loop corrections are realised with Higgs effective field theory (HEFT) assuming infinite top-quark mass.
The generated events are then corrected with a generator level bin-by-bin reweighting of the $m_{HH}$ distribution,
which is calculated with finite top-quark mass in full NLO corrections~\cite{Borowka:2016ehy}.
The branching fractions of the Higgs boson are assumed to be equal to the SM predictions~\cite{deFlorian:2016spz}.
For the SM $HH$ search, upper limits are extracted for the cross-section $\sigma_{\rm ggF} (pp \to HH)$
of $HH$ production and are normalised by the SM $HH$
cross-section $\sigma^{\rm SM}_{\rm ggF} (pp \to HH)$. The limits are
determined assuming that all kinematic properties of the $HH$ pair are
those predicted by the SM, particularly the $m_{HH}$ distribution, and only
the total ggF production cross-section, $\sigma_{\rm ggF} (pp \to HH)$, is allowed to deviate from its
SM value. The theoretical uncertainties of $\sigma_{\rm ggF}^{\rm SM}$
are less than 10\%~\cite{deFlorian:2016spz} and are not included in
the fit results.
 
\begin{figure}[tbp]
\centering
\includegraphics[width=0.65\textwidth]{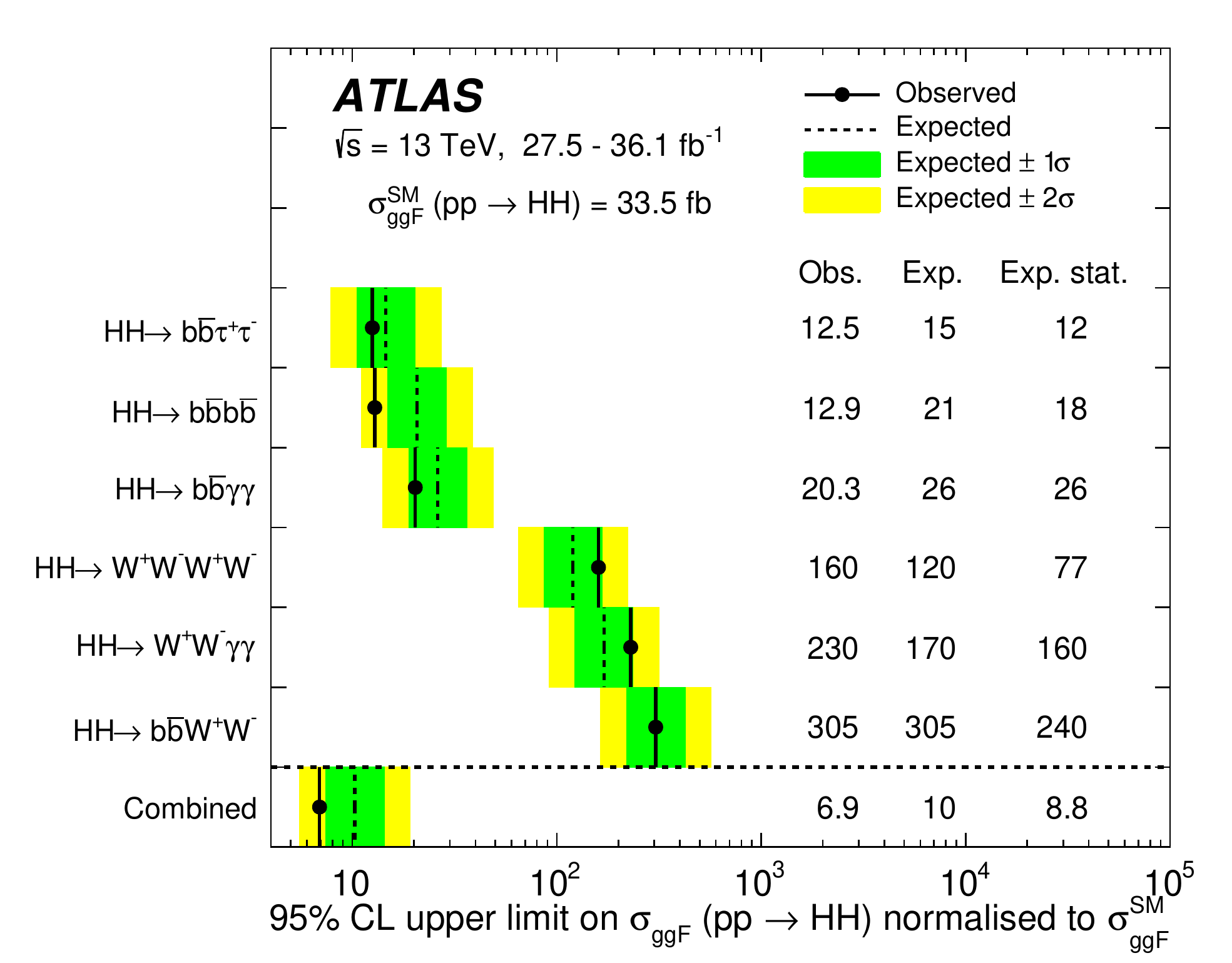}
\caption{Upper limits at 95\% CL on the cross-section of the ggF
SM $HH$ production normalised to its SM
expectation $\sigma^{\rm SM}_{\rm ggF} (pp \to HH)$ from the
\bbtautau, \bbbb, \bbyy, \wwww, \wwyy and \bbww searches, and their statistical
combination. The column ``Obs.'' lists the observed limits, ``Exp.'' the expected limits with all statistical and
systematic uncertainties, and ``Exp. stat.'' the expected limits
obtained including only  statistical uncertainties in the fit.}
\label{fig:NRuplim}
\end{figure}

The upper limits at 95\% CL on the cross-section of the ggF
Higgs boson pair production normalised to $\sigma_{\rm
ggF}^{\rm SM}$ are shown in Figure \ref{fig:NRuplim} for the
individual final states and their combination.
The upper limit for each final state is obtained from a fit with minimal
changes from previously published results.
The changes include an update of the ggF  Higgs boson pair production cross-section
from 33.4 fb to 33.5 fb for all final states. Additionally, the \bbtt
final state included theoretical uncertainties on the ggF inclusive cross-section, $\sigma_{\rm ggF}^{\rm SM}$,
which are not considered  in the present treatment, and the  \bbyy
final state is updated to use an asymptotic approximation
to calculate the observed limit instead of the pseudo-experiment method used for its publication. This results in a 10\%
change in the observed limit of \bbyy. Moreover, the impact of the
asymptotic approximation on all final states combined is found  to be 5\%.
 
The combined observed (expected) upper limit on the SM
$HH$ production is $\lorSMObsFull$~($\lorSMExpFull$) $\times$ $\sigma^{\rm SM}_{\rm ggF} (pp \to HH)$.
The expected limit is  similar  to the CMS result
of 12.8 $\times$ $\sigma^{\rm SM}_{\rm ggF} (pp \to HH)$.
The observed limit is more stringent for the ATLAS result than the CMS result
of 22.2 $\times$ $\sigma^{\rm SM}_{\rm ggF} (pp \to HH)$
because the three leading channels (\bbbb, \bbtt and \bbyy) have a data deficit
in ATLAS and an excess in CMS~\cite{Sirunyan:2018two}, remaining
however within the two 2$\sigma$ uncertainty interval.
Detailed comparisons can be found in Ref.~\cite{DiMicco:2019ngk}.
 
The impact of the systematic uncertainties has been evaluated by
recomputing the limit without their inclusion. The limit is then reduced
by 13\% when removing all of them.
The main sources of systematic uncertainty are the modelling of the backgrounds,
the statistical uncertainty of simulated events
and the $\tau$-lepton reconstruction and identification. When removed
the limit reduces by  5\%, 3\% and 2\%, respectively.
 
\section{Constraints on the Higgs boson self-coupling}
The results in Figure \ref{fig:NRuplim} show that the sensitivity of
the SM $HH$ search is driven by the final states \bbbb,
\bbtt and \bbyy. These final states are used to set constraints on
the Higgs boson self-coupling modifier $\kappa_{\lambda} =
\lambda_{HHH}/\lambda_{HHH}^{\rm SM}$.
After setting all couplings to fermions and bosons to their SM values, a scan of the
self-coupling modifier $\kappa_{\lambda}$ is performed. The
$\kappa_{\lambda}$ factor affects both the production cross-section and
the kinematic distributions of the Higgs boson pairs, by modifying the
$\mathcal{A}_2$ production amplitude. It can also affect the Higgs
boson branching fractions due to NLO electroweak corrections
\cite{Degrassi:2016wml}, but this dependence is neglected in the following.
 
The signal used in the $\kappa_{\lambda}$ fit was simulated
according to the following procedure. For each value of $\kappa_{\lambda}$  the $m_{HH}$
spectrum is computed at the generator-level, using the leading-order (LO) version of
{\sc MadGraph5\_aMC@NLO}~\cite{Alwall:2014hca} with the NNPDF 2.3 LO~\cite{Ball:2012cx}
PDF set, together with {\sc Pythia 8.2}~\cite{Sjostrand:2014zea}
for the showering model using the {\sc A14} tune~\cite{ATL-PHYS-PUB-2014-021}.  Because only one amplitude of Higgs boson pair production depends on
$\kappa_{\lambda}$, linear combinations of three LO samples generated
with different values of $\kappa_{\lambda}$  are sufficient to make predictions for any value of $\kappa_{\lambda}$.
Binned ratios of the $m_{HH}$ distributions to the SM distribution are computed for all  $\kappa_{\lambda}$ values
and then used to reweight the
events of NLO SM $HH$ signal samples, generated using the full
detector simulation.
This procedure is validated
by comparing kinematic distributions obtained with the
reweighting procedure applied to the LO SM sample
and  LO samples generated with the actual $\kappa_{\lambda}$ values set in the event generator.
The two sets of distributions are found to be in agreement.
This procedure assumes that higher order QCD corrections on the differential cross-section as a function of $m_{HH}$ are independent of $\kappa_{\lambda}$.
The reweighted NLO signal sample
is used to compute the signal
acceptance and the kinematic distributions for different values of
$\kappa_{\lambda}$.
 
This letter presents $\kappa_{\lambda}$ results for the first time
in the ATLAS \bbbb and \bbtt final states and incorporates the previously published result for
the \bbyy final state. The $\kappa_{\lambda}$ analyses closely follow  the SM $HH$ search, with some
exceptions which are discussed below for each final state.
 
\begin{figure}[htbp]
\begin{center}
\subfloat[]{\includegraphics[width=0.42\textwidth]{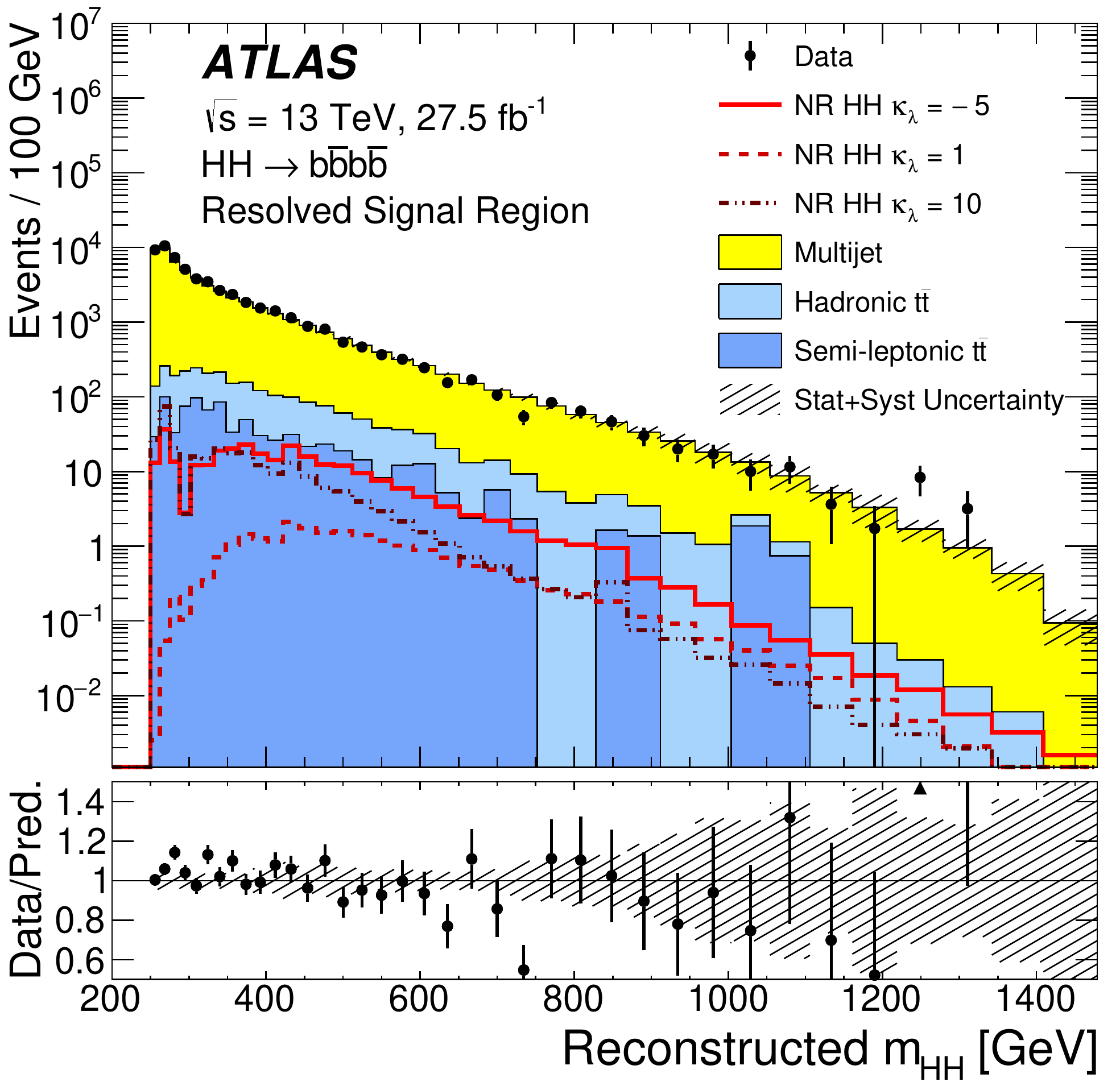}\label{fig:m4j_4b}} \\
\subfloat[]{\includegraphics[width=0.45\textwidth]{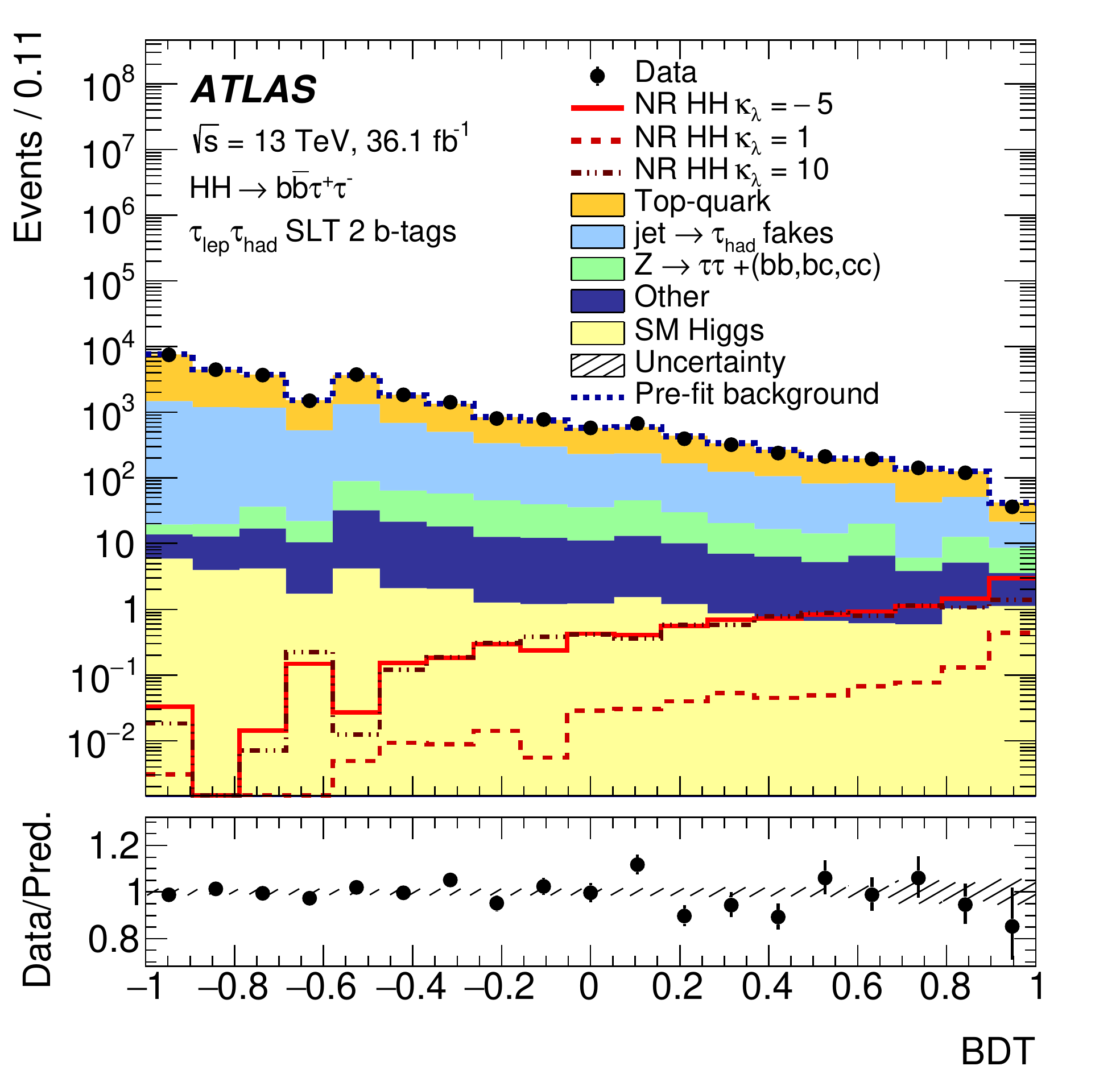}\label{fig:BDT_bbttLepHad}}
\subfloat[]{\includegraphics[width=0.45\textwidth]{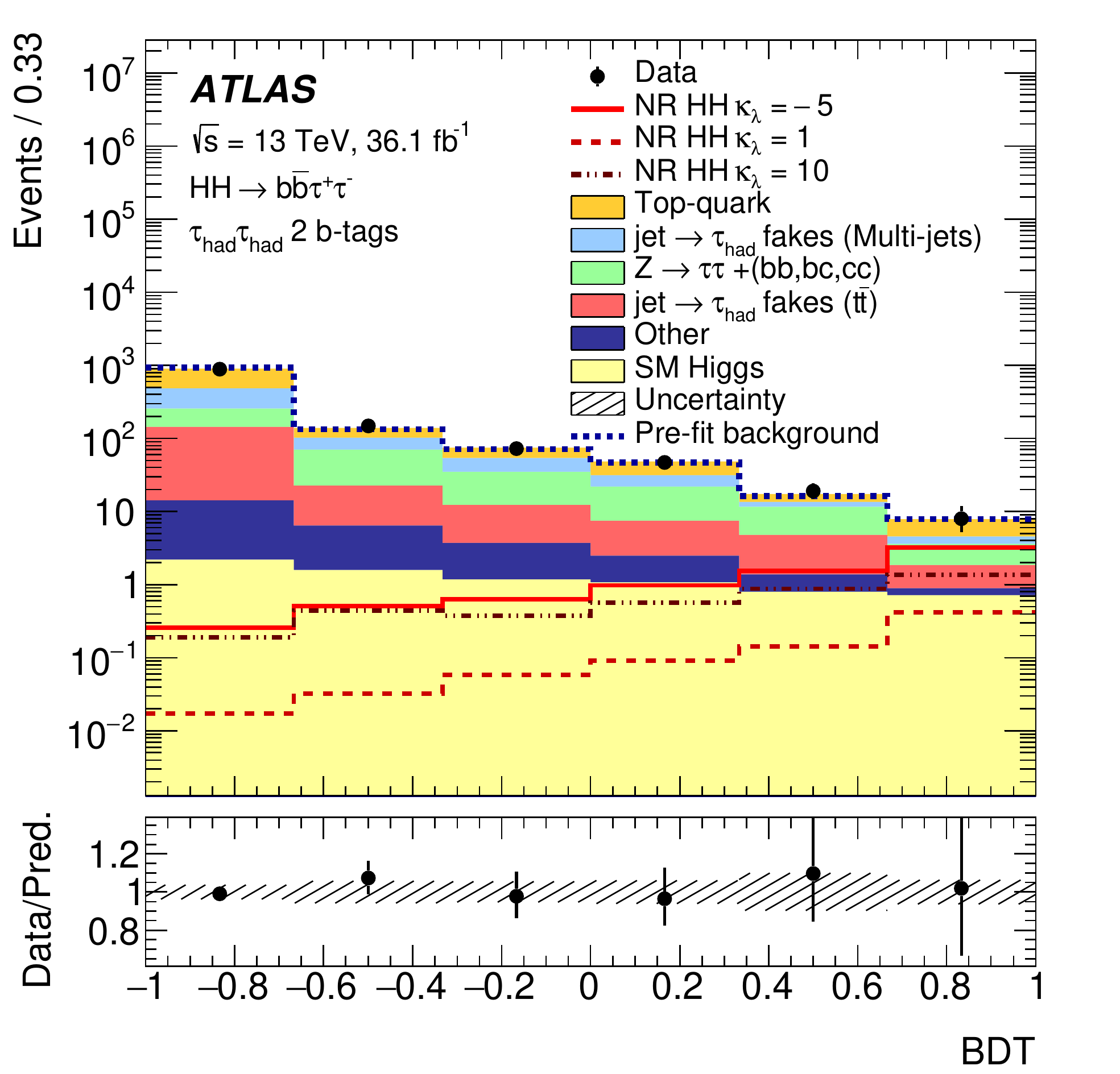}\label{fig:BDT_bbttHadHad}}
\end{center}
\caption{Final discriminants used in the $\kappa_{\lambda}$-scan
analysis for the \bbbb and the \bbtt final states.
(a) shows the reconstructed $m_{HH}$ distribution in the \bbbb
analysis; backgrounds include data-driven multi-jet processes (Multijet),
$t\bar{t} \to W^+W^-b\bar{b}$ with both $W$ bosons decaying hadronically (Hadronic
$t\bar{t}$) and $t\bar{t} \to W^+W^-b\bar{b}$ with at least one of the $W$ bosons decaying leptonically
(Semileptonic $t\bar{t}$).
(b) and (c) show the BDT distributions in the \bbtt analysis for the
$\taulep\tauhad$ and the $\tauhad\tauhad$ channels, respectively. The main backgrounds are
$t\bar{t}$ and single-top-quark production (Top-quark), the background
arising from jets faking hadronic $\tau$-lepton decays (${\rm jet} \to \tau_{\rm had}$
fakes), $Z\to \tautau$ plus two heavy-flavour jets [${\rm Z} \to \tau\tau+(bb,bc,cc)$],
SM single Higgs boson production (SM Higgs) and other minor
backgrounds (Other). The shaded area includes the systematic uncertainty of
the total background expectation due to the statistics of simulated events and
all experimental and theoretical systematic uncertainties.
In Figures (b) and (c) the  uncertainty band is not shown in the
upper panes because it is too small to be seen.
The signal distribution is overlaid for $\kappa_{\lambda} =
-5, 1, 10$  and is normalised to its expected yield.
}  \label{fig:final_discriminants}
\end{figure}

\begin{figure}[htbp]
\centering
\subfloat[]{\includegraphics[width=0.434\textwidth]{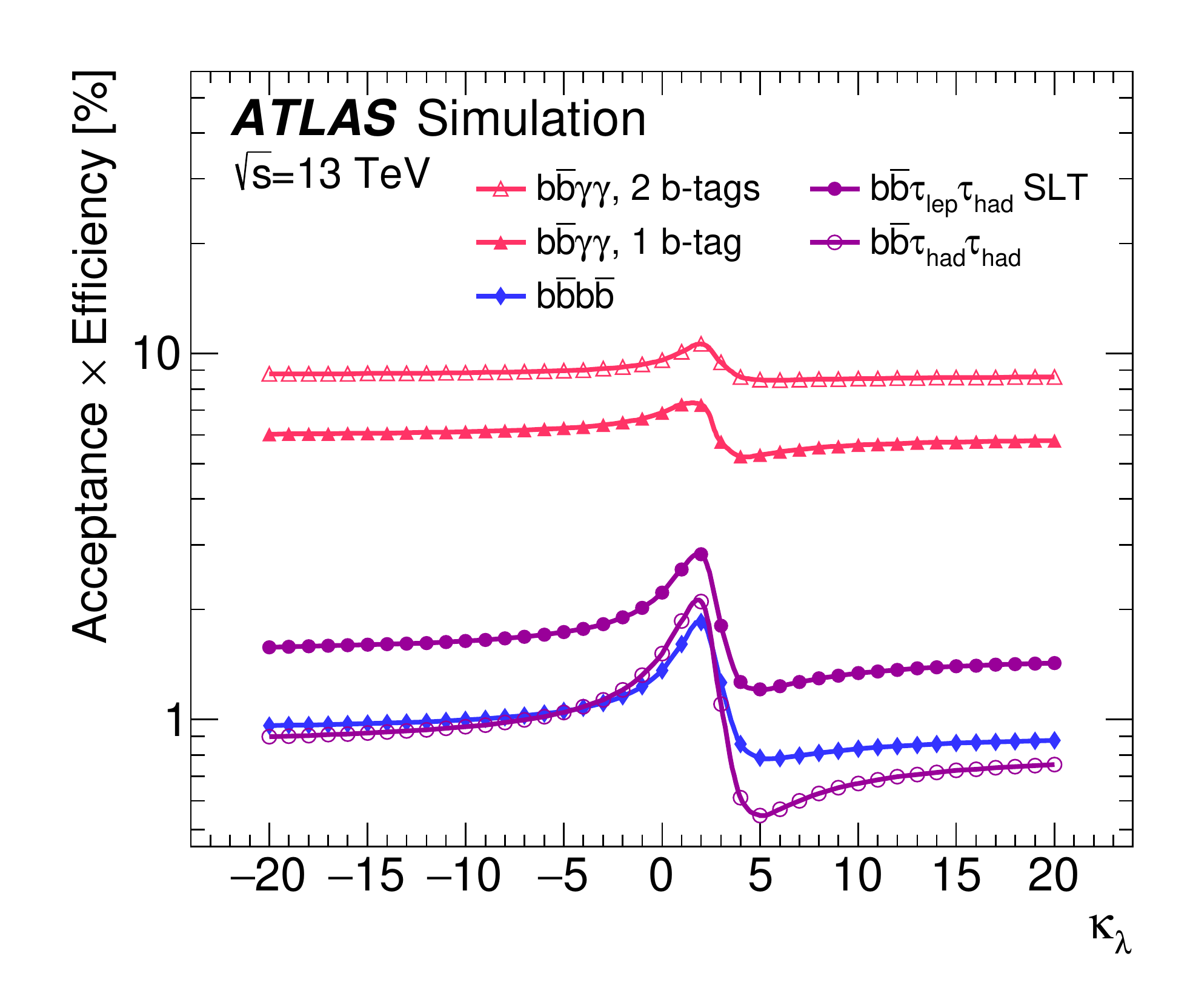}\label{fig:klacc}}
\subfloat[]{\includegraphics[width=0.566\textwidth]{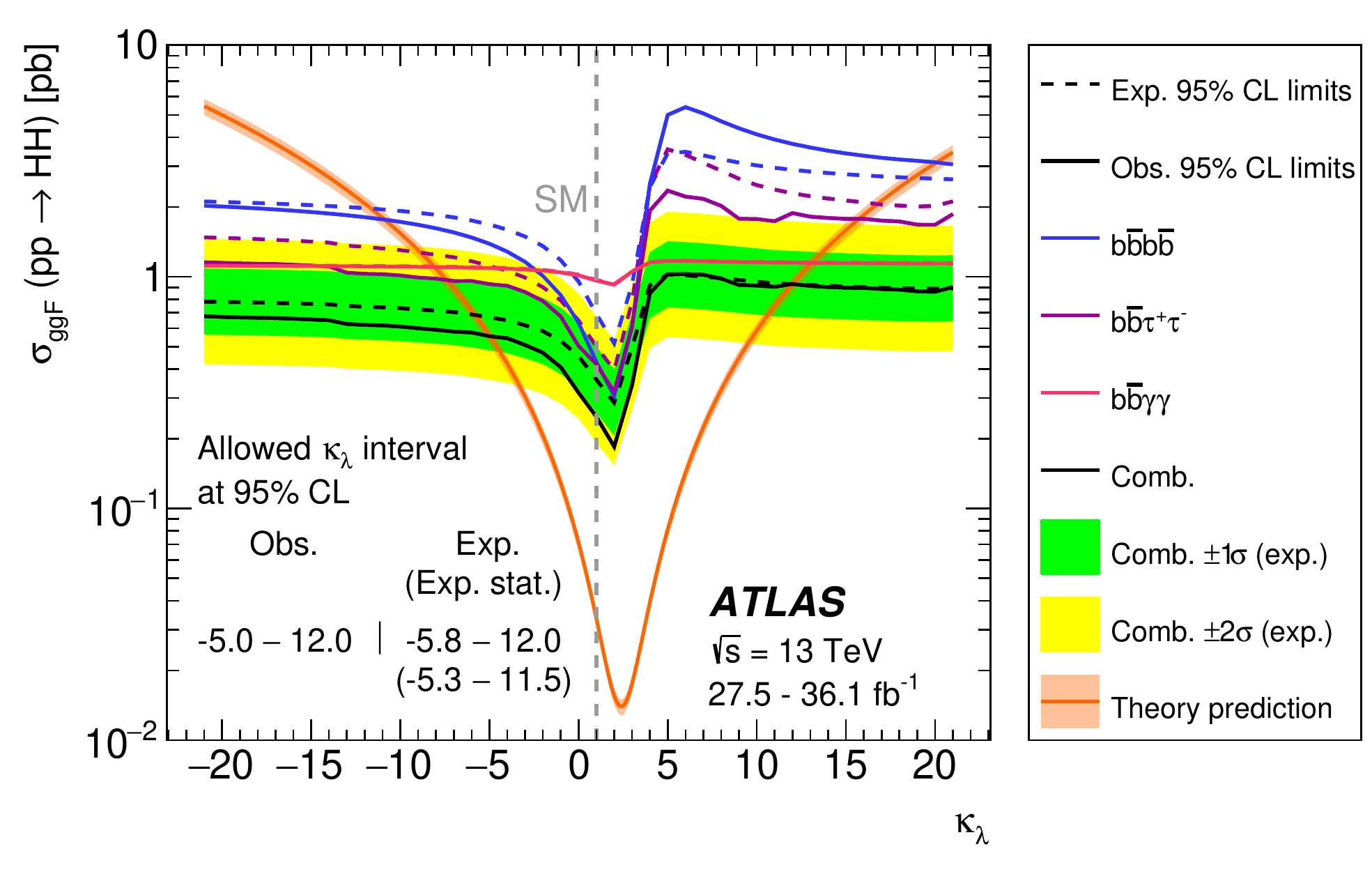}\label{fig:limit-lambda}}
\caption{(a) Signal acceptance times efficiency as a function of
$\kappa_{\lambda}$ for the \bbbb, \bbtt and \bbyy analyses. The
\bbbb curve is the average of the 2015 and 2016 curves weighted by
the integrated luminosities of the two datasets. (b) Upper limits
at 95\% CL on the cross-section of the ggF non-resonant SM $HH$ production
as a function of $\kappa_{\lambda}$.
The observed (expected) limits are shown as solid (dashed) lines. In
the \bbyy final state,
the observed and expected limits coincide. The $\pm1\sigma$ and $\pm2\sigma$ bands are only shown for
the combined expected limit. The theoretical prediction of the cross-section as a function of
$\kappa_{\lambda}$ is also shown.
The effect of non-SM Higgs decay branching fractions due to $\kappa_{\lambda}$ variations
is not taken into account, which impacts the $\kappa_{\lambda}$ intervals by no more than 7\%.}
\end{figure}

\begin{itemize}
\item In the \bbbb final state,
the same analysis selection and final discriminant are used in the $\kappa_{\lambda}$-scan
analysis and in the SM $HH$ search. The distribution of the final
discriminant $m_{HH}$ is shown in Figure~\ref{fig:m4j_4b},
where, with the exception of a small excess in the region below
300~\GeV~\cite{Aaboud:2018knk} and a small deficit in the 500-600~\GeV region, good agreement between data and the expected background is observed.
The shape of the $m_{HH}$ distribution has a strong dependence on
$\kappa_{\lambda}$, and the signal acceptance varies by
a factor 2.5 over the probed range of  $\kappa_{\lambda}$-values  ($-20 \le \kappa_{\lambda} \le 20$) shown
in Figure~\ref{fig:klacc}. The two effects together determine how the
exclusion limits on the cross-section
of the $HH$ production vary as a function of $\kappa_{\lambda}$.
\item In the \bbtt final state, as in the SM $HH$
search, both $\tau_{\rm lep}\tau_{\rm had}$ and $\tau_{\rm
had}\tau_{\rm  had}$ events are used. In contrast
with the SM $HH$ search, LTT
$\tau_{\rm lep}\tau_{\rm had}$ events (see Section~\ref{ana_descr})
are not used given their negligible contribution.
The SM $HH$ search and the $\kappa_{\lambda}$-scan analysis use
the same sets of variables to build  BDT discriminants.
For the $\kappa_{\lambda}$-scan the BDTs are retrained using the NLO
SM signal sample reweighted
with $\kappa_{\lambda}=20$, ensuring good sensitivity over the whole range of probed $\kappa_{\lambda}$-values.
The BDT score distributions are used in the fit to compute the
final results. The shape of the \bbtt BDT distributions does not show
a  $\kappa_{\lambda}$ dependence as strong as in the \bbbb final state, as can be seen in Figure~\ref{fig:final_discriminants}.
The sensitivity of this analysis is instead affected by a variation  in the
signal acceptance by up to a factor of three over the probed range of $\kappa_{\lambda}$-values, as shown in Figure~\ref{fig:klacc}.
\item In the \bbyy final state, the loose selection is used in the
$\kappa_{\lambda}$-scan analysis because the average transverse momentum of the Higgs bosons is lower
at large values of $\kappa_{\lambda}$, where  $|\mathcal{A}_2|^2$
dominates the production cross-section.
As in the SM $HH$ search, the statistical analysis is
performed using the $m_{\gamma\gamma}$ distribution, which
does not depend on $\kappa_{\lambda}$. The signal
acceptance varies by about 30\% over the probed
range of $\kappa_{\lambda}$-values, as shown in
Figure~\ref{fig:klacc}. In the previously published analysis
\cite{Aaboud:2018ftw}, LO samples were used for the computation of the
signal acceptance, while in this paper the NLO reweighted samples
are used, as described above.
\end{itemize}

\floatsetup[figure]{style=plain,subcapbesideposition=center}
 
The signal acceptance times efficiency as a function of $\kappa_{\lambda}$ is
shown in Figure~\ref{fig:klacc}. Given that, for each final state,  the same selection was applied over the full scanned $\kappa_{\lambda}$ range,
the shape of the acceptance times efficiency curve is determined by the variation
of the event kinematics as a function of $\kappa_{\lambda}$.
For high values of $|\kappa_{\lambda}|$ the $\mathcal{A}_2$ term
dominates the total amplitude, causing a softer $m_{\rm HH}$ spectrum,
and thus a lower acceptance times efficiency.
Around $\kappa_{\lambda}$ = 2.4 the interference
between $\mathcal{A}_1$ and $\mathcal{A}_2$ amplitudes is maximal,
producing the hardest $m_{\rm HH}$ spectrum and, consequently, the highest signal acceptance times efficiency.
 
In each  analysis, and in their combination,
the 95\% CL upper limit on the $\sigma_{\rm ggF} (pp \to HH)$ cross-section is computed for different values of $\kappa_{\lambda}$.
The  results  are shown in Figure~\ref{fig:limit-lambda}.
The  theoretical $\sigma_{\rm ggF} (pp \to HH)$ cross-section as
a function of $\kappa_{\lambda}$ is overlaid in the figure.  It is computed
by multiplying  the $HH$ SM cross-section $\sigma^{\rm SM}_{\rm ggF} (pp \to HH)$
by the ratio $R\left (\kappa_{\lambda} \right )$ of the $pp \to HH$
cross-section computed at $\kappa_{\lambda}$, $\sigma_{\rm ggF}^{\kappa_{\lambda}}(pp \to HH)$ to the
same quantity $\sigma_{\rm ggF}^{\kappa_{\lambda}=1}(pp \to HH)$ computed at $\kappa_{\lambda}=1$.
The $R(\kappa_{\lambda})$ factor is computed at NNLO+NNLL with
the infinite top-quark mass approximation ~\cite{YR4Tab63}.
The resulting observed (expected) confidence interval at 95\% CL for $\kappa_{\lambda}$ is:
\limitLambdaOBSlow~$<\kappa_{\lambda}<$~\limitLambdaOBShigh~(\limitLambdaEXPlow~$<\kappa_{\lambda}<$~\limitLambdaEXPhigh)~.
 
In Figure~\ref{fig:limit-lambda} the shape of the upper-limit curves
approximately follows  the inverse of the signal acceptance shown in Figure~\ref{fig:klacc}.
In the \bbbb analysis, the observed limits are more stringent than the
expected limits at low values of $\kappa_{\lambda}$. For these $\kappa_{\lambda}$ values the signal  $m_{HH}$ distributions
have significant populations in the region 500-600~GeV, where the
data deficit sits, as explained above.
For larger values of $\kappa_{\lambda}$ the $m_{HH}$ distribution is shifted to lower $m_{HH}$ values, and thus
the excess in data below 300 GeV leads to the observed limits being less stringent than expected.
In the \bbtt final state the observed limits are more stringent than the expected
limits over the whole range of $\kappa_{\lambda}$,  due to a deficit of
data relative to the background predictions at high values of the
BDT score. The \bbyy limit shows a weaker dependence on $\kappa_{\lambda}$ than
the \bbbb and \bbtt limits because the \bbyy acceptance varies less
as function of $\kappa_{\lambda}$.

The 95\% CL allowed $\kappa_{\lambda}$ intervals are given in Table~\ref{tab:klambda_interval}.
The systematic uncertainties weaken the $\kappa_{\lambda}$
limits by less than 10\% relative to those obtained
with only statistical uncertainties. The final state least (most) affected by
systematic uncertainties is \bbyy (\bbbb).
The Higgs boson branching fraction depends on $\kappa_{\lambda}$
due to NLO electroweak corrections \cite{Degrassi:2016wml}.
This dependence is neglected in the present treatment, but its overall
impact on the allowed $\kappa_{\lambda}$ interval is evaluated to be
no more than 7\%. Theory uncertainties on the signal cross
section shown  in Fgure~\ref{fig:limit-lambda} are not taken into
account when computing the $\kappa_{\lambda}$ limits in
Table~\ref{tab:klambda_interval}, they affect the limit by less than
8\%.

\begin{table}[htbp]
\begin{center}
{\def\arraystretch{1.3}\tabcolsep=8pt
\begin{tabular}{l|r@{}c@{}l|r@{}c@{}l|r@{}c@{}l}
& \multicolumn{9}{c}{Allowed $\kappa_{\lambda}$ interval at 95\% CL} \\
\hline
Final state &\multicolumn{3}{c|}{Obs.}&\multicolumn{3}{c|}{Exp.}&\multicolumn{3}{c}{Exp. stat.}\\
\hline
\bbbb &$\lambbbbObsLoFull$&~~---~~&\lambbbbObsHiFull&\lambbbbExpLoFull&~---~~&\lambbbbExpHiFull&\lambbbbExpLoSTAT&~---~~&\lambbbbExpHiSTAT\\
\bbtt & \lambbttObsLoFull&~---~~&\lambbttObsHiFull&\lambbttExpLoFull&~---~~&\lambbttExpHiFull&\lambbttExpLoSTAT&~---~~&\lambbttExpHiSTAT\\
\bbyy &\lambbyyObsLoFull&~---~~&\lambbyyObsHiFull&\lambbyyExpLoFull&~---~~&\lambbyyExpHiFull&\lambbyyExpLoSTAT&~---~~&\lambbyyExpHiSTAT\\
\hline
Combination &\limitLambdaOBSlow&~---~~&\limitLambdaOBShigh&\limitLambdaEXPlow&~---~~&\limitLambdaEXPhigh&\limitLambdaEXPlowSTATONLY&~---~~&\limitLambdaEXPhighSTATONLY\\
\end{tabular}
}
\end{center}
\caption{Allowed  $\kappa_{\lambda}$ intervals at 95\% CL for the \bbbb, \bbtt and
\bbyy final states and their combination.  The column~``Obs.'' lists the observed results,~``Exp.''
the expected results obtained including all statistical   and
systematic uncertainties in the fit, and~``Exp. stat.'' the expected
results obtained including only the statistical uncertainties.
The effect of non-SM Higgs decay branching fractions due to $\kappa_{\lambda}$ variations
is not taken into account, which impacts the $\kappa_{\lambda}$ intervals by no more than 7\%.} \label{tab:klambda_interval}
\end{table}

\section{Combination of results for resonant Higgs boson pair production}
\label{sec:res}
 
The resonance decaying into a pair of Higgs bosons is assumed
to be either a heavy spin-0 scalar particle, $S$, with a narrow width or a spin-2 KK  graviton, $G_{\text{KK}}$.
 
The search for the heavy scalar particle $S$ is performed with all six
final states included in this combination. With the exception of \bbtautau and
\bbbb,   all signal samples were simulated at NLO with {\sc MadGraph5\_aMC@NLO} using
the CT10 PDF set. The matrix-element generator was interfaced to
\textsc{Herwig++} with the \textsc{UE-EE-5-CTEQ6L1}
tune. The \bbtautau final state uses an LO model generated with {\sc MadGraph5\_aMC@NLO} using the
NNPDF 2.3 LO PDF set interfaced to \textsc{Pythia 8.2} with the A14
tune, while the \bbbb final state uses the
same LO event generator but interfaced to \textsc{Herwig++}
with the \textsc{UE-EE-5-CTEQ6L1} tune.
 
\begin{figure}[htbp]
\centering
\subfloat[]{\includegraphics[width=0.50\textwidth]{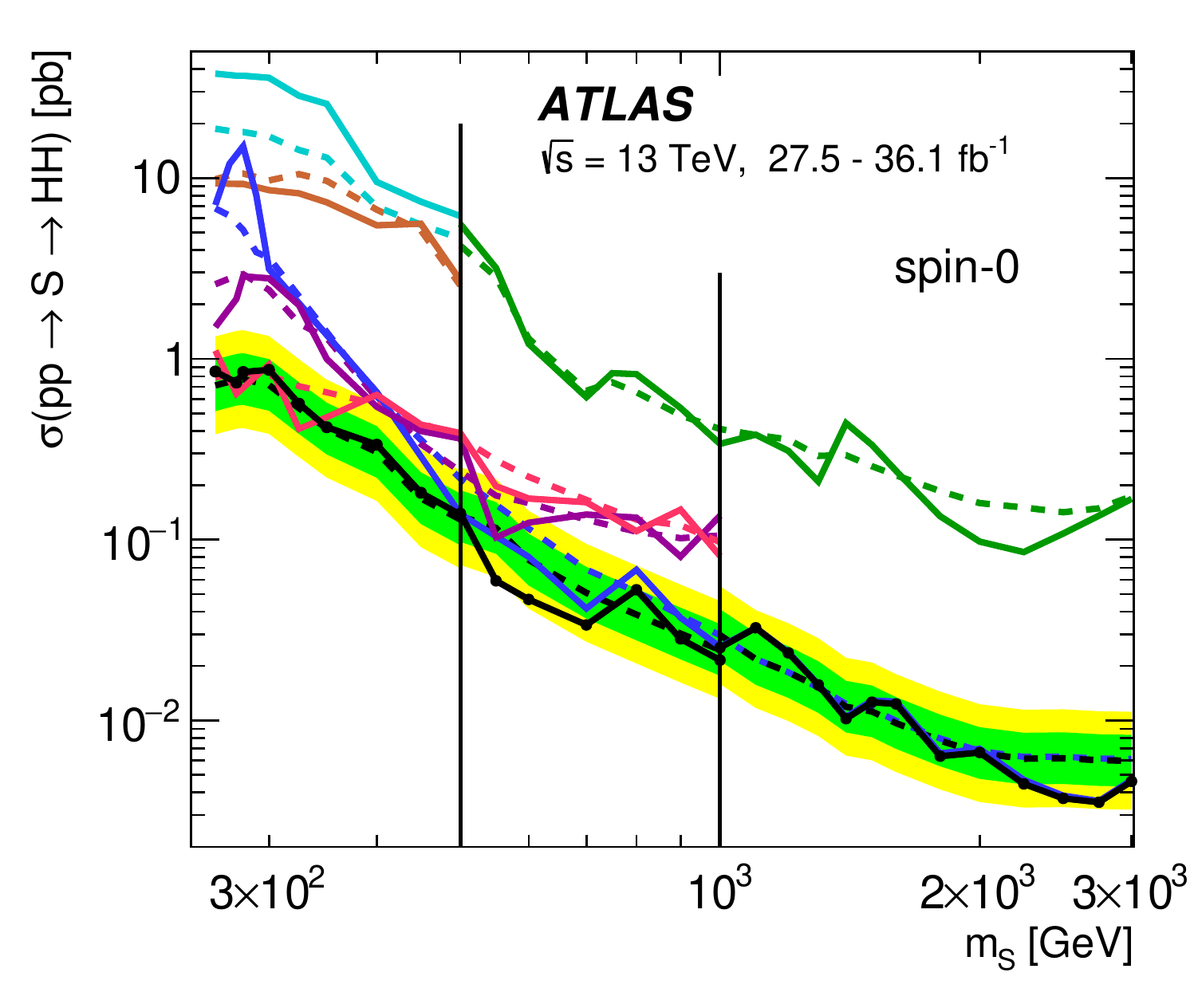}\label{fig:res-spin0}}
\quad ~~ \subfloat{\includegraphics[width=0.40\textwidth]{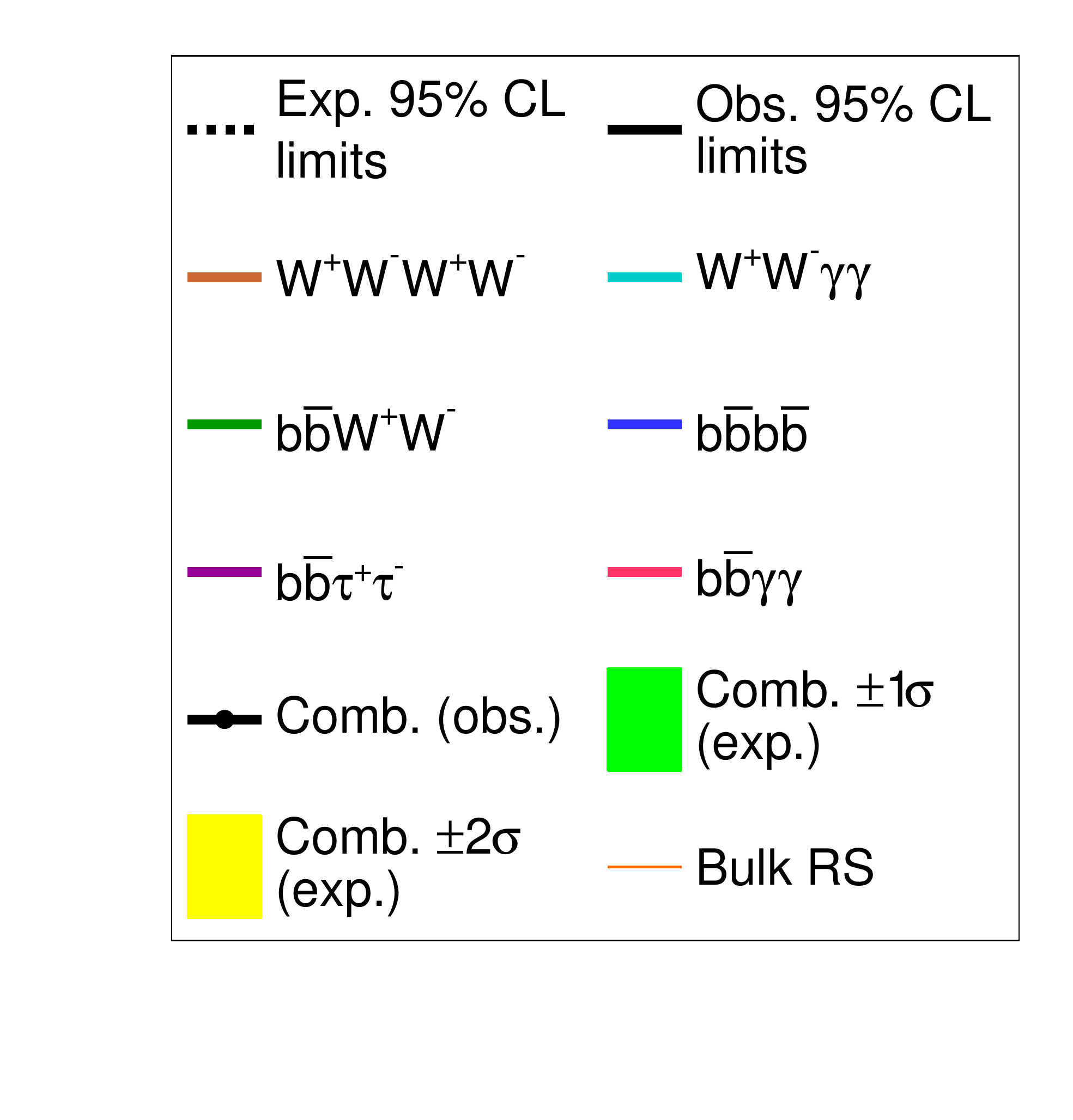}}
\quad \quad ~~ \\
\setcounter{subfigure}{1}
\subfloat[]{\includegraphics[width=0.50\textwidth]{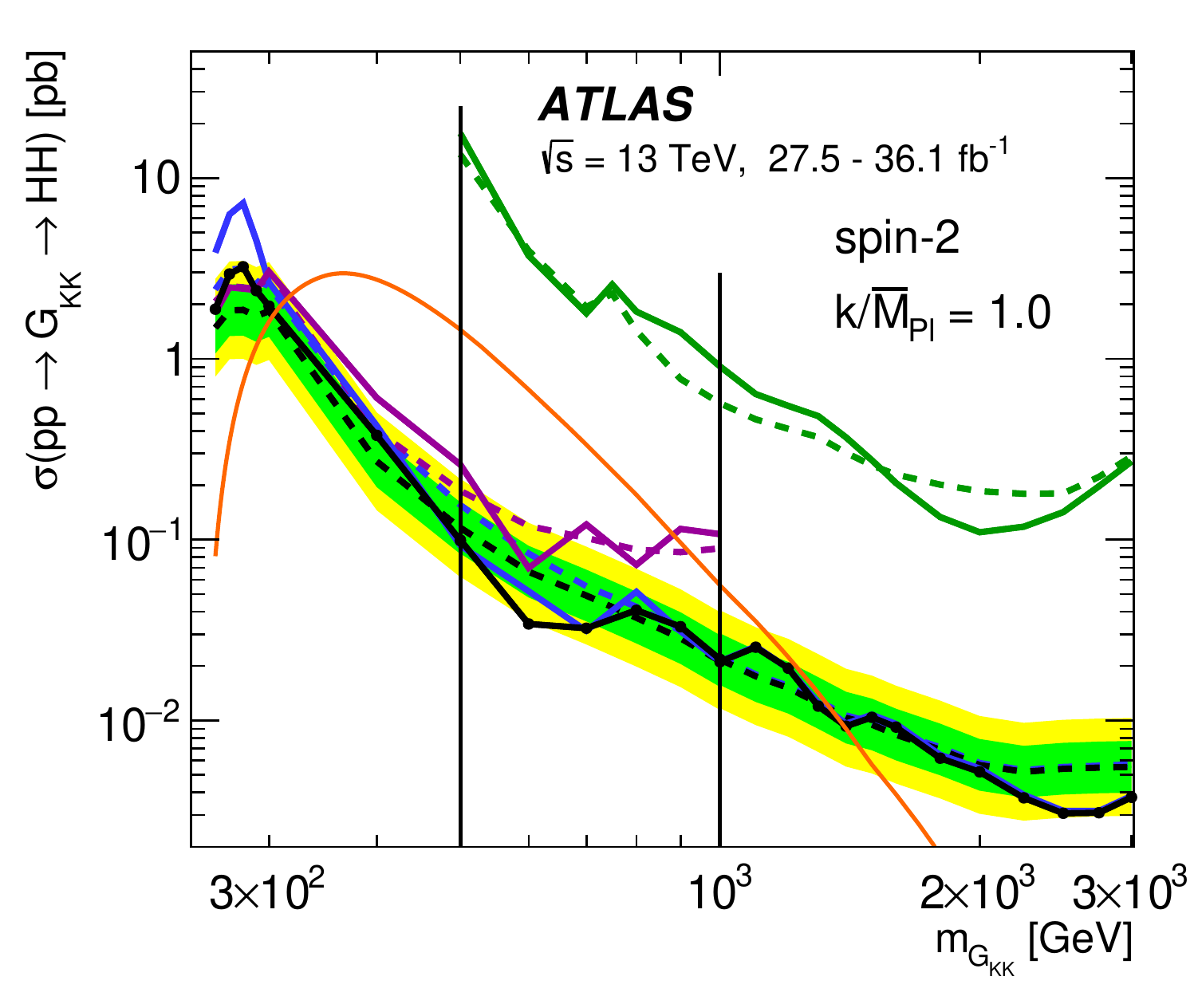}\label{fig:res-spin2-c1}}
\subfloat[]{\includegraphics[width=0.50\textwidth]{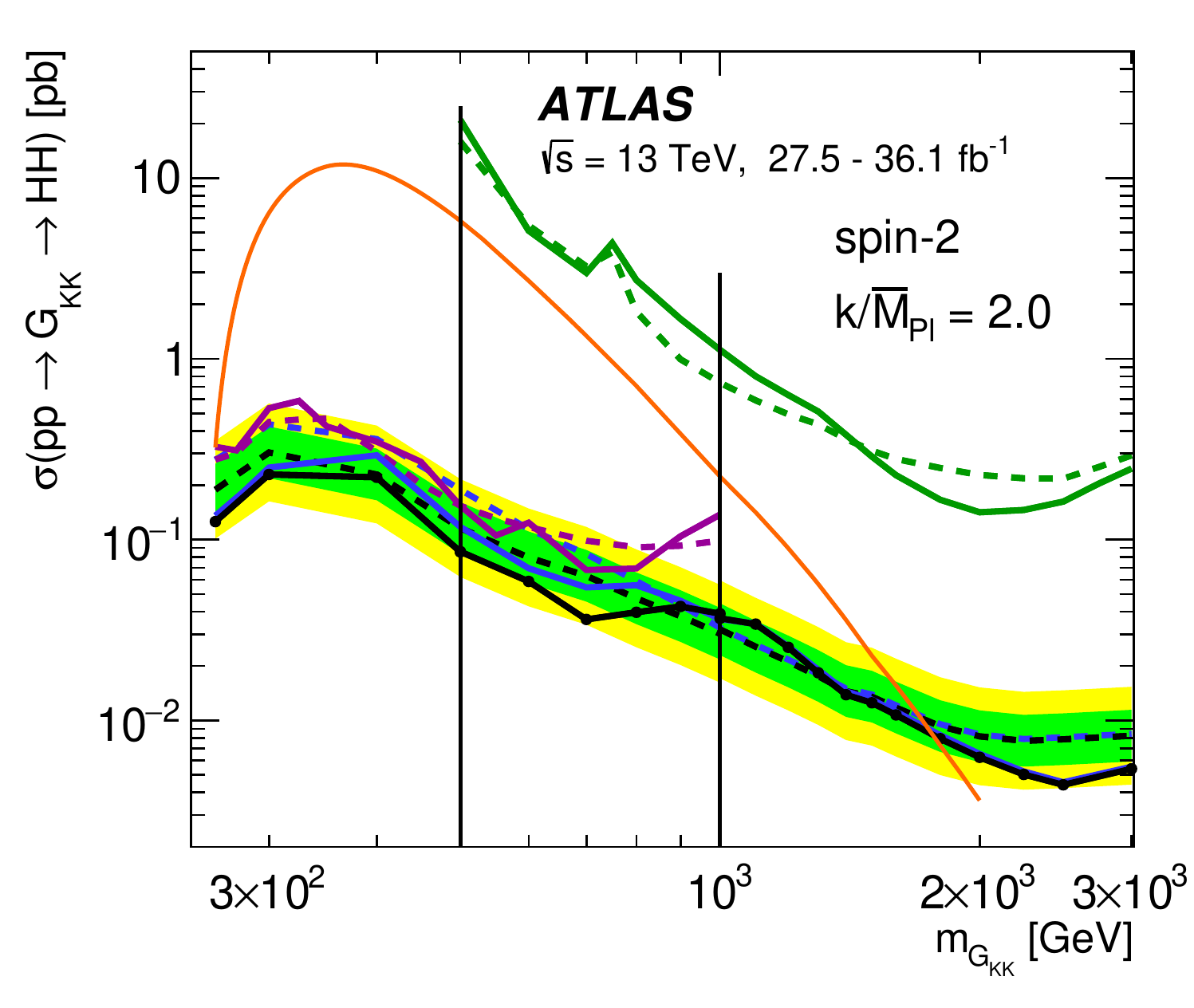}\label{fig:res-spin2-c2}}
 
\caption{Upper limits at 95\% CL on the cross-section of the resonant
Higgs boson pair production for (a) a spin-0 heavy scalar,
(b) a spin-2 KK graviton with $k/\overline{M}_{\text{Pl}} = 1$ and
(c) a spin-2 KK graviton with $k/\overline{M}_{\text{Pl}} = 2$.
The observed (expected) limits are shown as solid (dashed) lines.
The $\pm1\sigma$ and $\pm2\sigma$ bands are only shown for the expected limits of the combination.
Only the \bbbb, \bbww and \bbtt search results are used in the spin-2 resonant
combination.
The vertical black lines in each pannel indicate mass intervals
where different final states are combined.}
\label{fig:res-spin2}
\end{figure}

The scalar resonance search is performed in the mass range 260--3000
GeV, and within this range no statistically significant excess is
observed.  In the combination, the largest observed deviation from the
background expectation is 1$\sigma$ for the search mass range.
The combined upper limit on the cross-section is shown as a function of the resonance mass in Figure~\ref{fig:res-spin0}.
Systematic uncertainties have a sizeable effect on the upper limits
depending on the probed resonance mass.
The total impact of systematics or the impact of a
single systematic uncertainty has been evaluated by computing the
percentage reduction of the upper limit obtained by removing all
systematic uncertainties or a particular source.
Overall the systematic uncertaintes affect the limit  by 12\%
(11\%) for a resonance mass of 1~(3)~TeV.
Among them, the largest systematic uncertainties are due to the
modelling of the backgrounds, impacting the upper limit by 7\% (9\%) at 1~(3)~TeV.
The second leading systematic uncertainty comes from $b$-tagging, that
affects the upper limit by 2\% at 1~TeV, but  its impact  is negligible at 3~TeV
where relative background and statistical uncertainties increase significantly.
At 3~TeV the second leading systematic uncertainty is related to the
jet energy scale and resolution, changing the limit  by 2\%.
Interpretations in specific spin-0 BSM models are provided in Section \ref{sec:bsm}.
 
The search for a spin-2 KK graviton is
performed with the \bbbb, \bbww and \bbtt final states only.
Gravitons were simulated using an LO model in {\sc MadGraph5\_aMC@NLO}
with the NNPDF 2.3 LO PDF set interfaced to \textsc{Pythia 8.2} with the
A14 tune. The resonance width changes with the graviton mass
and depends on the parameter $k/\overline{M}_{\text{Pl}}$, where $k$ is the curvature of the warped extra dimension
in the bulk RS model and $\overline{M}_{\text{Pl}} = 2.4 \times 10^{18}~\GeV$ is the effective
four-dimensional Planck mass. The search is performed for models with
$k/\overline{M}_{\text{Pl}}$ equal to 1 and 2.
For $k/\overline{M}_{\text{Pl}}$ = 1 (2), the width ranges from 3\%
(11\%) for a 0.3 TeV graviton mass to 6\% (25\%) for a 3 TeV graviton mass.

The upper limits in the $G_{\text{KK}}$ search are shown as a function of the resonance mass in Figures~\ref{fig:res-spin2-c1}
and~\ref{fig:res-spin2-c2} for $k/\overline{M}_{\text{Pl}}$ equal to 1 and
2, respectively. In the combination, the largest observed deviation
from the background expectation is 1.5$\sigma$ (0.7$\sigma$)
for the search mass range with $k/\overline{M}_{\text{Pl}}=$1 (2).
Exclusion ranges on the $KK$ graviton mass are obtained by comparing the
upper limit with the production cross section calculated at LO.
In the case of $k/\overline{M}_{\text{Pl}} = 1$, the bulk RS model
is excluded at 95\% CL in the graviton mass range from \spinTwoCOneMassObsLow~GeV to \spinTwoCOneMassObsHigh~GeV.
In the case of $k/\overline{M}_{\text{Pl}} = 2$, the model is excluded at 95\% CL for graviton masses from
260~GeV, where the scan starts, to \spinTwoCTwoMassObsHigh~GeV.
 
The impact of the systematic uncertainties on the upper limits on $G_{\text{KK}}$
has a small dependence on the resonance mass. It is  $\sim$20\%
over the whole mass range  for $k/\overline{M}_{\text{Pl}}=1$, and 29\% (25\%) at a mass of 1~TeV (3~TeV) for $k/\overline{M}_{\text{Pl}}=2$.
The largest systematic uncertainties are from the modelling of the
backgrounds, affecting the limit by 11\% (15\%) at 1~TeV (3~TeV) for $k/\overline{M}_{\text{Pl}}=1$
and 16\% (21\%) at 1~TeV (3~TeV) for $k/\overline{M}_{\text{Pl}}=2$.
For $k/\overline{M}_{\text{Pl}}=1$, the subleading systematic uncertainties come from
$b$-tagging at low $G_{\text{KK}}$ mass, that affect the limit by 3\%,
and from jet energy scale and resolution at high mass, that affect the
upper limit by 2\% (3\%) at 1~TeV (3~TeV).
For $k/\overline{M}_{\text{Pl}}=2$, subleading systematic uncertainties are
from jet energy scale and resolution, impacting the upper limits by
5\%  at 1~TeV and 4\% at 3~TeV.
The systematic uncertainties affect upper limits more for $k/\overline{M}_{\text{Pl}}=2$
than for $k/\overline{M}_{\text{Pl}}=1$, because the natural width
of the signal graviton is four times larger with $k/\overline{M}_{\text{Pl}}=2$.

\clearpage
\section{Constraints on the hMSSM and EWK-singlet models}
\label{sec:bsm}
Exclusion limits are also presented for two specific models,
namely the EWK-singlet model~\cite{Hill:1987ea,Veltman:1989vw,Robens:2015gla,deFlorian:2016spz}
and the hMSSM model~\cite{Djouadi:2013vqa,Maiani:2013hud,Djouadi:2013uqa,Djouadi:2015jea,deFlorian:2016spz}.
The sensitivity of the \bbww, \wwww and \wwyy final states to these models
is negligible, so the presented results combine only the \bbbb, \bbtt
and \bbyy final states.
 
\begin{figure}[htbp]
\centering
\subfloat[]{\includegraphics[width=0.45\textwidth]{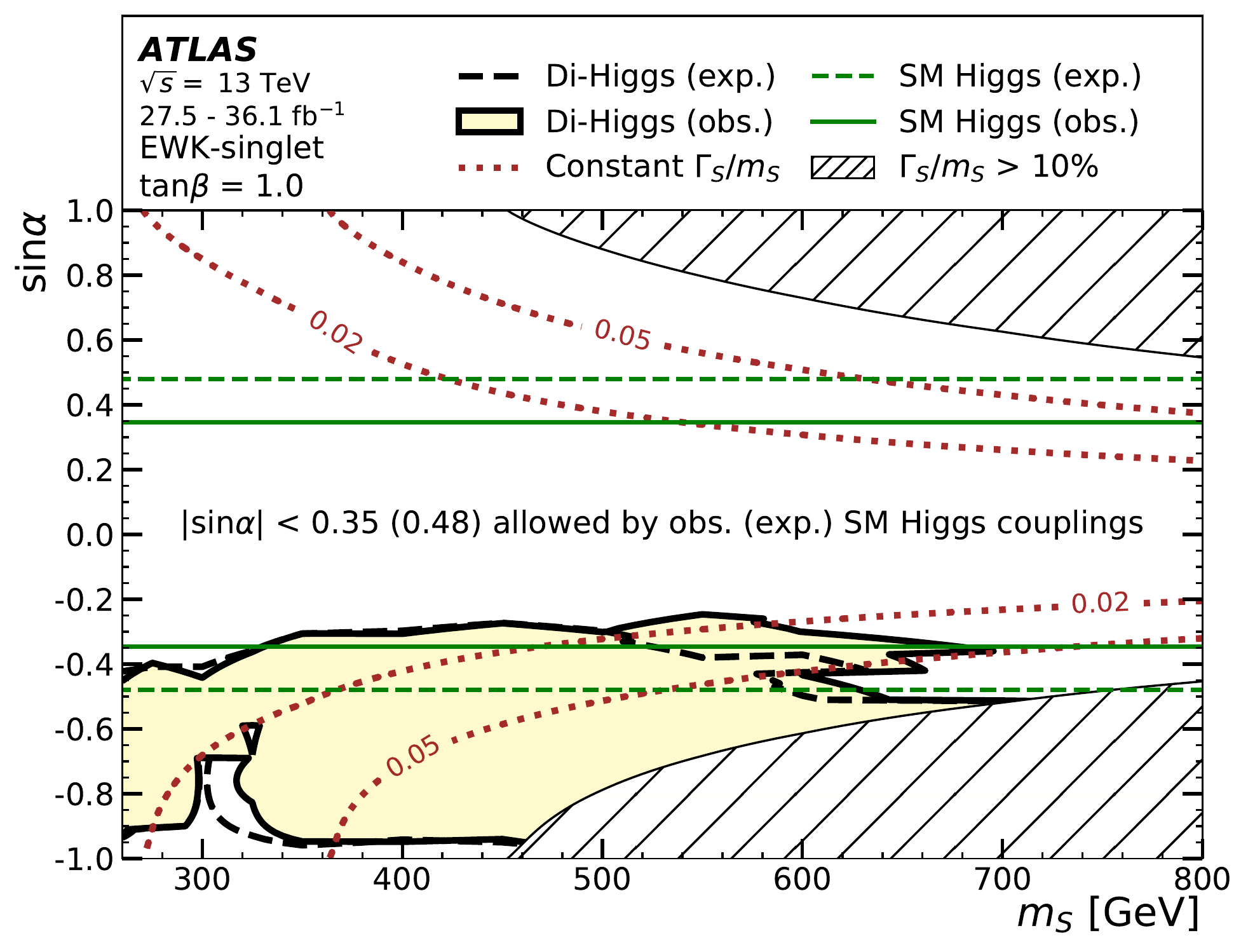}\label{fig:singletb1}}
\subfloat[]{\includegraphics[width=0.45\textwidth]{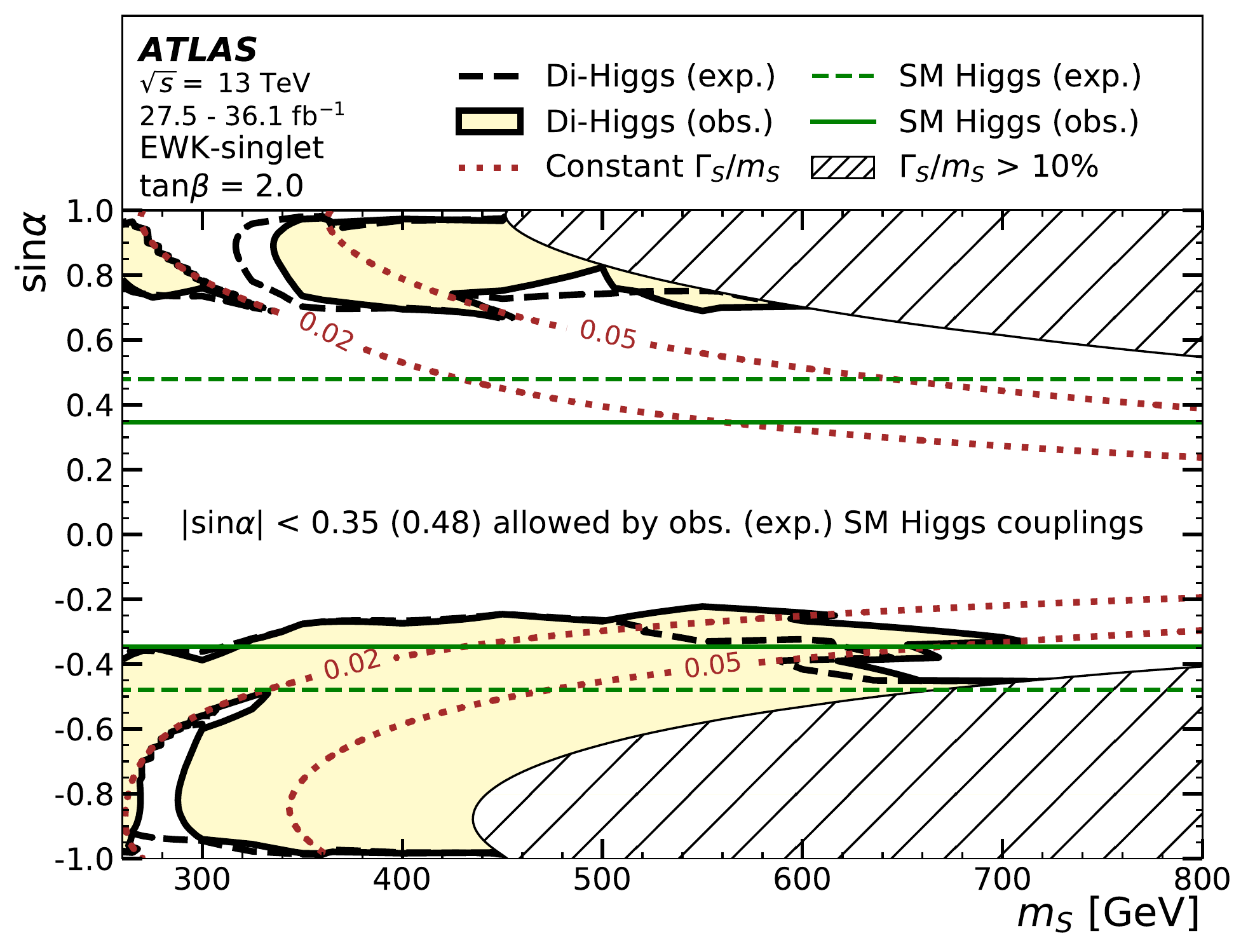}\label{fig:singletb2}}
\caption{Excluded regions for the EWK-singlet model for two values
of the tan $\beta$ parameter:   (a) tan $\beta$ = 1 and (b) tan
$\beta$ = 2. Indirect constraints from SM Higgs coupling measurements~\cite{Aad:2015pla}
are shown with horizontal lines. The
dotted lines indicate the separation  between regions where the
resonance width is larger than 2\% and 5\% of the resonance mass.  The hatch-marked area corresponds to
regions that cannot be excluded because the width of the resonance
exceeds 10\% of the resonance mass, corresponding to the maximum of
the experimental mass resolution among all analysed final states.}
\label{fig:bsm}
\end{figure}

\begin{figure}[htbp]
\centering
\subfloat[]{\includegraphics[width=0.45\textwidth]{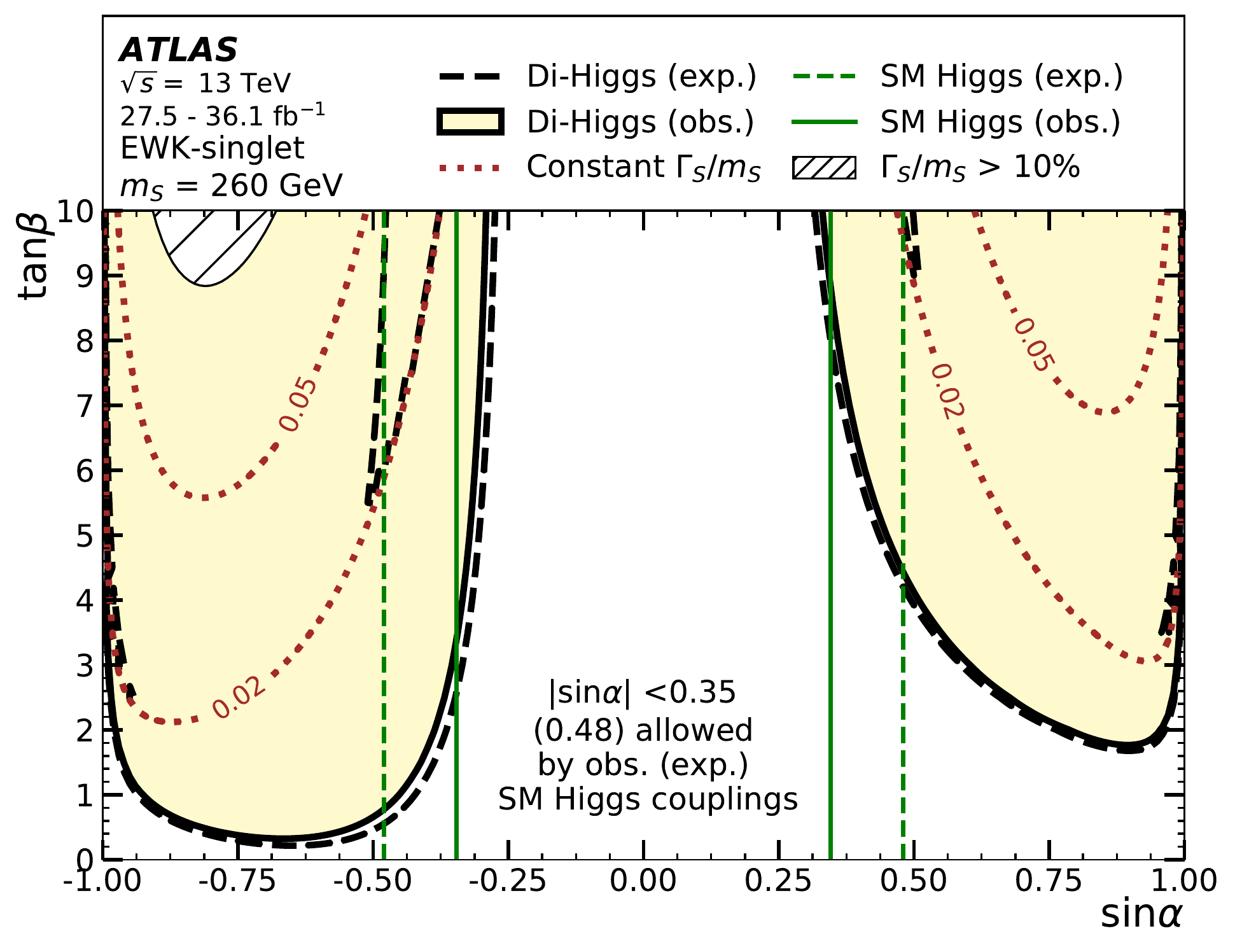}\label{fig:singlet_260}}
\subfloat[]{\includegraphics[width=0.45\textwidth]{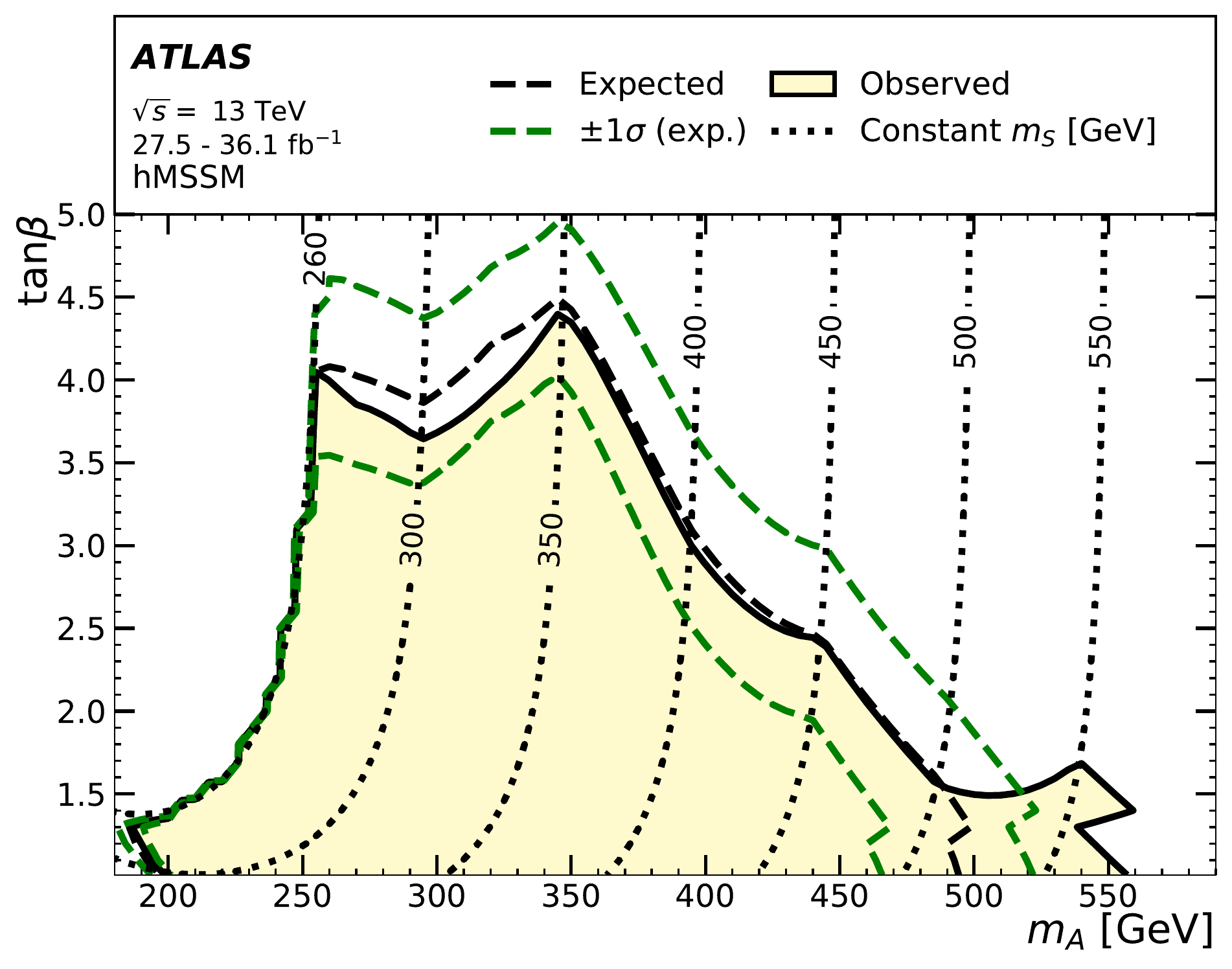}\label{fig:hmssm}}
 
\caption{Excluded regions in (a) the EWK-singlet model for $m_S=$ 260 GeV and (b) the hMSSM model
using the experimental upper limits obtained in the spin-0 resonance
searches. In the EWK-singlet exclusion, the indirect contraints from SM Higgs coupling measurements~\cite{Aad:2015pla}
are shown with vertical lines.}
\end{figure}

For the EWK-singlet model, the experimental limits on the spin-0
resonance (as reported in Section~\ref{sec:res}) are interpreted as constraints in
the $m_S$--$\sin\alpha$ plane (where $m_S$ is the resonance mass) for $\tan \beta =1$ and $\tan \beta =2$,
shown in Figure~\ref{fig:singletb1} and Figure~\ref{fig:singletb2}
respectively. The expected cross-section for each point in the parameter space is
obtained by scaling the heavy Higgs cross-section calculated at
NNLO+NNLL \cite{deFlorian:2016spz} with singlet coupling
modifiers. The branching fractions are computed with {\sc sHDECAY} \cite{Costa:2015llh}.
In this model, the width of the heavy scalar can be large in some regions of
the parameter space. Due to the use of narrow-width signal models in
the event generation, results
presented here are valid only in regions of the model parameter-space
where the resonance width ($\Gamma_S$) is smaller than the experimental
resolution at the resonance mass. This holds when $\Gamma_S/m_S < 2$\% for
\bbyy, $\Gamma_S/m_S<5$\% for \bbbb and $\Gamma_S/m_S<10$\% for \bbtt.
Therefore, the excluded region in
the plot is obtained by combining the three final states for $\Gamma_S/m_S <
2$\%, by combining the \bbbb and \bbtt final states for 2\% $<
\Gamma_S/m_S < 5$\%, and using only \bbtt for 5\% $< \Gamma_S/m_S < 10$\%. The hatched region
shows points where $\Gamma_S/m_S \ge 10$\%, where no exclusion can be
provided.
Figure \ref{fig:singlet_260} shows limits for the (sin $\alpha$,
tan $\beta$) parameter space for $m_S = 260$ GeV where, due to the limited
decay phase space, the
resonance width is narrow in a wide region of the  parameter space.

The experimental limits on a spin-0 resonance are also interpreted as constraints in
the $m_A$--$\tan\beta$ plane of the hMSSM model in
Figure~\ref{fig:hmssm}.
The expected cross-section for each point in the parameter space is
obtained using the gluon-gluon fusion cross-section from {\sc SUSHI
1.5.0} \cite{Harlander:2012pb,Harlander:2016hcx} and the branching
fractions computed with {\sc HDECAY 6.4.2} \cite{Djouadi:1997yw}.
 
The excluded region is more than doubled along $\tan\beta$ relative to the previous combined results
in Ref.~\cite{Aad:2015xja} at 8~TeV,  and excludes values of $m_A$ from 190 GeV to 560 GeV depending on $\tan\beta$.
The kink at low $\tan\beta$ and high $m_A$ values is caused by removing
the
\bbyy final state from the combination in the region where the predicted width of
the heavy CP-even Higgs boson is larger than the experimental resolution on
$m_S$ in the \bbyy analysis.
 
\clearpage
\section{Conclusion}
\label{sec:conclusion}
 
A statistical combination of six final states \bbbb, \bbww, \bbtautau, \wwww, \bbyy and
\wwyy, is presented for the search for non-resonant and resonant production of Higgs boson pairs.
These searches use up to \LUMI\ of proton--proton collision data at
13~\TeV\ recorded with the ATLAS detector at the LHC\footnote{All
results are available in digital format on HEPDATA at the following
link: \url{https://www.hepdata.net/record/90521}}.
 
In both resonant and non-resonant searches, no statistically  significant excess of events above
the Standard Model  predictions is found. For the Standard Model $HH$ production mode, the observed (expected)
95\% confidence level upper limit on the gluon--gluon fusion $pp \to HH$ cross-section
is \lorSMObsFull~(\lorSMExpFull) times the Standard Model prediction.
The expected limit is comparable to the CMS result,
while the observed limit is significantly stronger than CMS's
due to a data deficit compared to expected background in ATLAS and an excess in CMS.
For the resonant case, upper limits are set on the production
cross-section of heavy spin-0 and spin-2 resonances decaying into  pairs of Higgs bosons in the mass range
260--3000~\GeV.
 
Upper limits on the $pp \to HH$ cross-section are also computed as a function of the Higgs boson self-coupling
modifier $\kappa_{\lambda} = \lambda_{HHH}/\lambda_{HHH}^{\text{SM}}$, by
combining the \bbbb, \bbtautau and \bbyy final states.
The combination excludes $\kappa_{\lambda}$ values outside the range
$\limitLambdaOBSlow < \kappa_{\lambda} < \limitLambdaOBShigh$
($\limitLambdaEXPlow < \kappa_{\lambda} <\limitLambdaEXPhigh$) at 95\%
confidence level in observation (expectation). The three final states are also combined to constrain the
Electroweak Singlet Model in the ($m_S$, sin $\alpha$) and the  (sin $\alpha$, tan $\beta$)
parameter spaces and the habemus Minimal Supersymmetric Standard Model in the ($m_A$, tan $\beta$) parameter space.

\section*{Acknowledgements}
 
% The next lines are included from the .//acknowledgements/Acknowledgements.tex input file
 
We thank CERN for the very successful operation of the LHC, as well as the
support staff from our institutions without whom ATLAS could not be
operated efficiently.
 
We acknowledge the support of ANPCyT, Argentina; YerPhI, Armenia; ARC, Australia; BMWFW and FWF, Austria; ANAS, Azerbaijan; SSTC, Belarus; CNPq and FAPESP, Brazil; NSERC, NRC and CFI, Canada; CERN; CONICYT, Chile; CAS, MOST and NSFC, China; COLCIENCIAS, Colombia; MSMT CR, MPO CR and VSC CR, Czech Republic; DNRF and DNSRC, Denmark; IN2P3-CNRS, CEA-DRF/IRFU, France; SRNSFG, Georgia; BMBF, HGF, and MPG, Germany; GSRT, Greece; RGC, Hong Kong SAR, China; ISF and Benoziyo Center, Israel; INFN, Italy; MEXT and JSPS, Japan; CNRST, Morocco; NWO, Netherlands; RCN, Norway; MNiSW and NCN, Poland; FCT, Portugal; MNE/IFA, Romania; MES of Russia and NRC KI, Russian Federation; JINR; MESTD, Serbia; MSSR, Slovakia; ARRS and MIZ\v{S}, Slovenia; DST/NRF, South Africa; MINECO, Spain; SRC and Wallenberg Foundation, Sweden; SERI, SNSF and Cantons of Bern and Geneva, Switzerland; MOST, Taiwan; TAEK, Turkey; STFC, United Kingdom; DOE and NSF, United States of America. In addition, individual groups and members have received support from BCKDF, CANARIE, CRC and Compute Canada, Canada; COST, ERC, ERDF, Horizon 2020, and Marie Sk{\l}odowska-Curie Actions, European Union; Investissements d' Avenir Labex and Idex, ANR, France; DFG and AvH Foundation, Germany; Herakleitos, Thales and Aristeia programmes co-financed by EU-ESF and the Greek NSRF, Greece; BSF-NSF and GIF, Israel; CERCA Programme Generalitat de Catalunya, Spain; The Royal Society and Leverhulme Trust, United Kingdom.
 
The crucial computing support from all WLCG partners is acknowledged gratefully, in particular from CERN, the ATLAS Tier-1 facilities at TRIUMF (Canada), NDGF (Denmark, Norway, Sweden), CC-IN2P3 (France), KIT/GridKA (Germany), INFN-CNAF (Italy), NL-T1 (Netherlands), PIC (Spain), ASGC (Taiwan), RAL (UK) and BNL (USA), the Tier-2 facilities worldwide and large non-WLCG resource providers. Major contributors of computing resources are listed in Ref.~\cite{ATL-GEN-PUB-2016-002}.
 
% End of text imported from the .//acknowledgements/Acknowledgements.tex input file

\printbibliography

\clearpage % ATLAS Collaboration author list
% Reference date of HDBS-2018-58 is 2019-04-02
% Author list last updated on date 29-OCT-19
% Data extracted on 29-Oct-2019 for paper reference HDBS-2018-58
% at 5:32pm
 
\begin{flushleft}
{\Large The ATLAS Collaboration}

\bigskip

G.~Aad$^\textrm{\scriptsize 101}$,    
B.~Abbott$^\textrm{\scriptsize 128}$,    
D.C.~Abbott$^\textrm{\scriptsize 102}$,    
A.~Abed~Abud$^\textrm{\scriptsize 70a,70b}$,    
K.~Abeling$^\textrm{\scriptsize 53}$,    
D.K.~Abhayasinghe$^\textrm{\scriptsize 93}$,    
S.H.~Abidi$^\textrm{\scriptsize 167}$,    
O.S.~AbouZeid$^\textrm{\scriptsize 40}$,    
N.L.~Abraham$^\textrm{\scriptsize 156}$,    
H.~Abramowicz$^\textrm{\scriptsize 161}$
% ...
\bigskip
\\

% ...
$^{40}$Niels Bohr Institute, University of Copenhagen, Copenhagen; Denmark.\\
% ...
$^{53}$II. Physikalisches Institut, Georg-August-Universit\"{a}t G\"ottingen, G\"ottingen; Germany.\\
% ...
$^{70}$$^{(a)}$INFN Sezione di Pavia;$^{(b)}$Dipartimento di Fisica, Universit\`a di Pavia, Pavia; Italy.\\
% ...
$^{93}$Department of Physics, Royal Holloway University of London, Egham; United Kingdom.\\
% ...
$^{101}$CPPM, Aix-Marseille Universit\'e, CNRS/IN2P3, Marseille; France.\\
$^{102}$Department of Physics, University of Massachusetts, Amherst MA; United States of America.\\
% ...
$^{128}$Homer L. Dodge Department of Physics and Astronomy, University of Oklahoma, Norman OK; United States of America.\\
% ...
$^{156}$Department of Physics and Astronomy, University of Sussex, Brighton; United Kingdom.\\
% ...
$^{161}$Raymond and Beverly Sackler School of Physics and Astronomy, Tel Aviv University, Tel Aviv; Israel.\\
% ...
$^{167}$Department of Physics, University of Toronto, Toronto ON; Canada.\\
% ...
$^{a}$ Also at Borough of Manhattan Community College, City University of New York, New York NY; United States of America.\\
$^{b}$ Also at CERN, Geneva; Switzerland.\\
% ...

\end{flushleft}

% Created with Glance <Atlas.Glance@cern.ch>

\end{document}